\documentclass[prd,aps,showpacs,epsf]{revtex4} 
\usepackage{epsfig}
\usepackage{amsmath}
\usepackage{amssymb}

\begin{document}
\newcommand{\hf} {{1\over2}}
\def\eq#1{(\ref{#1})}
\def\cd#1{{\cal D}[#1]}
\def\ord#1{{\cal O}(#1)}
\def\tA{{\tilde A}}
\def\tG{{\tilde G}}
\def\tbe{{\tilde \beta}}
\def\tu{{\tilde u}}
\def\mb#1{{\vec{#1}}} 
\def\bbox#1{\mbox{\boldmath$#1$}}
\def\tr{\mbox{Tr}}
\def\mr#1{\mathrm{#1}}

\title{Renormalization-Group Analysis of the Generalized sine-Gordon Model\\
and of the Coulomb Gas for $d\ge 3$ Dimensions}

\author{I. N\'andori$^{1,2,3}$}

\author{U. D. Jentschura$^{1}$}

\author{K. Sailer$^2$}

\author{G. Soff$^1$}

\affiliation{$^1$ Institut f\"ur Theoretische Physik, 
Technische Universit\"at Dresden, \/\/
01062 Dresden, Germany}

\affiliation{$^2$ Department of Theoretical Physics, 
University of Debrecen, \/\/ 
H-4032, Debrecen, Hungary}

\affiliation{$^3$ Institute of Nuclear Research, P.O.Box 51,
H-4001 Debrecen, Hungary}

\date{\today}

\begin{abstract}
Renormalization-group (RG) flow equations have been derived 
for the generalized sine-Gordon model (GSGM) and the Coulomb gas (CG)
in $d~\geq~3$ of dimensions by means of Wegner's and Houghton's, and 
by way of the real-space RG approaches. The UV scaling laws determined 
by the leading-order terms of the flow equations are in qualitative 
agreement for all dimensions $d \geq 3$, independent of the 
dimensionality, and in sharp contrast to the special case $d=2$. 
For the $4$-dimensional GSGM it is demonstrated explicitly (by 
numerical calculations), that the blocked potential tends to a constant 
effective potential in the infrared (IR) limit, satisfying 
the requirements of periodicity and convexity. The comparison of 
the RG flows for the three-dimensional GSGM, the CG, and  the 
vortex-loop gas reveals a significant dependence on the 
renormalization schemes and the approximations used.
\end{abstract}


\maketitle

%
%
\section{Introduction} 

The renormalization of a two-dimensional one-component scalar field theory 
with a periodic self-interaction, and the two-dimensional generalized 
sine-Gordon model (GSGM) have been investigated for $d=2$ dimensions 
in great detail over the last three 
decades~\cite{kt,col,jkkn,wieg,samu,agg,Nienhuis,kerson,
creswick,gulacsi,wet,per,coulomb,jpg,fertig}.
Here, we only mention some of the main results:
\begin{itemize}
\item
The blocked potential for the two-dimensional GSGM tends to a constant
effective potential in the IR limit, independent of the field 
\cite{per,coulomb,jpg}. 
\item 
Following the procedure proposed in \cite{deconf}, one can identically  
rewrite the partition function of the $d$-dimensional GSGM in the form 
of a $d$-dimensional static Coulomb gas, which we call here
the equivalent Coulomb gas (ECG). For $d=2$ it is well-known that the 
GSGM and the ECG, as well as the XY model 
describing classical planar spins 
belong to the same universality class \cite{gulacsi}.
The XY model has a dual theory in the sense of \cite{savit} that
can be reformulated as a gas of topological excitations:
the two-dimensional vortex gas (VG) for $d=2$, and the 
three-dimensional vortex-loop gas (VLG) for $d=3$ (see \cite{gulacsi} and 
Refs. therein). For dimensions $d=2$, it was shown that the ECG and the 
VG can be transformed into one another by an appropriate duality 
transformation \cite{jkkn,Nienhuis,kerson,fertig} that inverts the 
temperature. Two-dimensional generalized models are well known where 
both the ECG and the VG are included as particular limiting cases 
\cite{jkkn,Nienhuis,kerson,fertig} and are self-dual 
under the duality transformation. 
\end{itemize}

While the relations of the various models for two dimensions are
well-established, those of the corresponding models for $d=3$ are not 
completely settled (see e.g. \cite{gulacsi}). In particular, it is not 
proven that the VLG and the three-dimensional XY-model belong to the same 
class of universality. Therefore, the investigation of the renormalization 
of the GSGM and the ECG for $d\ge 3$ dimensions is essential for a  
further clarification of this issue. The purpose of the present work
is threefold:
\begin{enumerate}
\item to investigate the renormalization of the GSGM for dimensions
$d\ge 2$ by the Wegner-Houghton renormalization-group (WH-RG) method
\cite{wh}
and to demonstrate numerically that the blocked potential for the
$4$-dimensional GSGM tends to a constant effective potential in the IR 
limit; and
\item to perform an RG analysis for the ECG for dimensions $d\ge 2$
by the real-space renormalization group (RS-RG) method using the
dilute gas approximation (DGA) \cite{Nienhuis}; and finally 
\item to compare the RG flows obtained for the GSGM, ECG 
and VLG \cite{nelson,shenoy} in the three-dimensional case.
\end{enumerate}
 
Renormalization of the  GSGM and that of the related models in 
$d\ge 3$ dimensions are of particular interest with regard 
to the numerous physical applications. In high-energy physics the haaron 
model \cite{deconf} of confinement is in need of a 
clarification of the IR behaviour of the GSGM in $d=4$ dimensions. 
In low-temperature physics the 
superfluid transition of He$^4$ belongs to the universality class of 
the three-dimensional XY model and it provides an exceptional opportunity 
for an experimental test of the RG predictions \cite{spaceshuttle}. 
In superconductivity the three-dimensional XY model (isotropic and 
anisotropic) was used to study the phase transitions in the 
three-dimensional flux-line lattice \cite{choi,shenoy2}. The three-dimensional 
XY model can also help us in understanding the flux motions in
superconductors with extremely high anisotropy; these have been observed
in recent experiments \cite{lsc}.

Due to the fact that the  GSGM and the ECG are equivalent at the level 
of the partition function, their  renormalization-group flows 
obtained by applying different 
RG methods are alternative approaches to tackle 
essentially the same problem. 
On the one hand, the WH-RG applied to the GSGM keeps the periodicity 
of the self-interaction potential in each infinitesimal blocking step. 
There is a price to pay: 
one has to handle infinitely many vertices 
in the process of searching the blocked action 
in the subspace of functionals periodic 
in the field variable. Furthermore, the requirements of periodicity and 
convexity \cite{conv} are laid upon the effective potential of the model 
which constrain it to a field-independent constant \cite{per}. The WH-RG 
has the advantage that (i) it can be extrapolated to the IR limit 
and (ii) is able to treat the spinodal instability \cite{tree}. Its main
disadvantage is that the truncation of the gradient expansion is unavoidable
and that the implementation of the sharp momentum cut-off 
can induce unphysical features of the RG flow going beyond the 
local-potential approximation (LPA)
\cite{morris}. By restricting ourselves to LPA, we ignore the 
wave-function renormalization, which is equivalent 
to neglecting the scale-dependence of the temperature beyond the 
tree-level and its physical consequences. 

On the other hand, the RS-RG applied to the ECG keeps intact the CG structure in 
each blocking step and therefore also the periodicity, albeit indirectly. 
It implements a sharp cut-off in real space and has 
been proven very adequate for applications to the two-dimensional Coulomb 
gas \cite{Nienhuis} and to the three-dimensional vortex-loop gas  
\cite{nelson,shenoy}. Nevertheless, it has the drawbacks that (i) no 
systematic approach is known to go beyond the 
dilute-gas approximation (DGA), and (ii)  the sharp 
cut-off in real space, which is effectively a smooth momentum cut-off, 
is inadequate to treat the 
spinodal instability. Because our WH-RG approach will show that the latter 
always occurs, the interpretation of the 
extrapolation of our RS-RG results for the ECG to the 
IR regime is problematic. Nevertheless, it is interesting to compare
the RS-RG results for the ECG in the UV regime
to those for the gas of topological excitations (e.g. the VLG
for $d=3$).

The phase transition of systems in the universality class of the 
two-dimensional Coulomb gas, known as the Kosterlitz-Thouless-Berezinskii 
phase transition, is a topological phase transition because  there is no 
spatial long-range order for $d=2$ dimensions at finite temperature  
due to the Mermin-Wagner theorem \cite{mermin}. For $d\ge3$ dimensions  
the Mermin-Wagner theorem does not hold any more. 
Therefore, one expects a modification of the phase structure of the
Coulomb gas in the number of dimensions $d>2$ as compared to the case with
$d=2$. As shown below, the UV scaling laws in $d>2$ for both the GSGM and 
the ECG share the property of a lacking Coleman fixed point, 
which by contrast is present for $d=2$ \cite{col}.

In  Sect. \ref{related} we introduce our notations and summarize RG flow 
equations obtained in the literature for the various models discussed
in the next sections. The RG flow and the blocked potential 
for the $4$-dimensional GSGM are determined numerically in 
Sect. \ref{flowGSGM} by the WH-RG method. 
In Sect. \ref{flowCG} the RG  flow obtained by the RS-RG approach for
the $d$-dimensional Coulomb gas is discussed. In Sect. \ref{duality} 
the RG flow of the three-dimensional Coulomb gas (ECG) is compared to the 
RG flow obtained in for the three-dimensional GSGM, and to those for the 
three-dimensional VLG \cite{shenoy}. Finally, the results are summarized
in Sect. \ref{summar}. RS-RG flow equations for the $d$-dimensional Coulomb 
gas are
derived in App. \ref{renecg}.

%
%
\section{GSGM and related models}\label{related}

\subsection{GSGM and ECG}\label{DCG}

First, we remind the reader that the partition function for the GSGM
can be identically rewritten in the form of the partition function
for a CG, that we shall call the equivalent CG (ECG). In its lattice 
regulated form the partition function of the GSGM is given as
\begin{equation}\label{ZGSGM1}
  Z_{\rm GSGM} = \int\cd{\phi} 
    \exp\left[- \hf a^{2d} \sum_{x,y} \phi_x d_{x,y} \phi_y -
      a^d \sum_x \sum_{n=1}^\infty u_n \cos (n\beta \phi_x) \right],
\end{equation}
where $a$ denotes the lattice spacing, and 
$ d_{x,y} = 2 a^{-d-2}  ( d\delta_{x,y} 
             - \sum_{\mu=1}^d\delta_{y,x+n_\mu a} ) $
is the inverse propagator. Performing the transformation of the field 
variable $\varphi_x=\beta \phi_x$, one finds
\begin{equation}\label{ZGSGM2}
  Z = \int\cd{\varphi} \exp\left[- { a^{2d}\over 2\beta^2}\sum_{x,y} 
      \varphi_x d_{x,y} \varphi_y -
      a^d \sum_x  \sum_{n=1}^\infty u_n \cos (n \varphi_x) \right].
\end{equation}
This partition function  can be identically rewritten in the form of 
the macrocanonical partition function for a CG applying the procedure 
proposed in \cite{deconf}:
\begin{enumerate}
\item One expands the exponential factor of the integrand with the 
periodic potential in  Taylor series, expresses $\cos (n\varphi)$ in 
terms of the exponential function, and introduces the integer valued 
variable, the charge $\sigma=\pm 1,\pm 2,\ldots$; the result of 
these operations is
\begin{equation}
Z= \left( \prod_x \int d\varphi_x \right)
 \sum_{\nu =0}^\infty
{(-1)^\nu\over 2^\nu \nu!}a^{\nu d} 
\sum_{\begin{array}{c} \scriptstyle x_1,x_2,\ldots,x_\nu\\[-0.5ex]
\scriptstyle \sigma_1,\ldots,\sigma_\nu\not=0\end{array}}\,
u_{\sigma_1}\cdots u_{\sigma_\nu}
\exp\left[ -{ a^{2d}\over 2\beta^2} \sum_{x,y}   \varphi_x d_{x,y} \varphi_y
 + i a^d   \sum_x  \rho_x \varphi_x \right],
\end{equation}
where the charge density, 
$   \rho_x = \sum_{j=1}^\nu \sigma_j a^{-d} \delta_{x,x_j} $,
has been introduced and $u_{-\sigma}=u_{\sigma}$ holds due to the $Z(2)$
symmetry of the potential.
\item Performing the Gaussian path integral, one obtains
\begin{subequations}
\label{ZCG}
\begin{eqnarray}
Z &=& \sum_{\nu =0}^\infty
{(-1)^\nu\over 2^\nu \nu!}a^{\nu d} 
\sum_{\begin{array}{c} \scriptstyle x_1,x_2,\ldots,x_\nu\\[-0.5ex]
\scriptstyle \sigma_1,\ldots,\sigma_\nu\not=0\end{array}}\,
u_{\sigma_1}\cdots u_{\sigma_\nu} \, I^{\sigma_1,\ldots ,\sigma_\nu}_{
x_1,\ldots, x_\nu}\,,\\[2ex]
I^{\sigma_1,\ldots ,\sigma_\nu}_{x_1,\ldots, x_\nu}
  &=& \exp\left[- \hf \mr{Tr} \ln G^{-1} - \hf \beta^2 
      \sum_{i,j=1}^\nu \sigma_i\sigma_j G_{x_i,x_j}\right],
\end{eqnarray}
\end{subequations}
with the (dimensionful) propagator
$    G_{x,y}=  a^{-2d} (d^{-1})_{x,y}$.
\end{enumerate}
Including the self-interaction terms into the (dimensionless) fugacities,
$   w_\sigma= -\hf a^d u_\sigma \exp[-\hf \sigma^2 \beta^2 G_{x_i,x_i} ]$
and replacing $G_{x,x}$  by the value $G(a)$ of the propagator at the 
cut-off distance $a$, the partition function \eq{ZCG} takes the form
\begin{eqnarray}\label{zcg}
&&   Z_{\rm ECG} \equiv  \sum_{\nu =0}^\infty Z^{[\nu]}=
 \sum_{\nu =0}^\infty
{1\over  \nu!} \int_{D_0}{d^d x_1\over a^d} \int_{D_1}{d^d x_2\over a^d}
\cdots \int_{D_{\nu-1}}
{d^d x_\nu\over a^d}
 \sum_{\sigma_1,\ldots,\sigma_\nu\not=0}
w_{\sigma_1}\cdots w_{\sigma_\nu} J^{\sigma_1,\ldots ,\sigma_\nu}_{
x_1,\ldots, x_\nu} ,
\end{eqnarray}
where
\begin{subequations}
\label{wtu}
\begin{eqnarray}
    J^{\sigma_1,\ldots ,\sigma_\nu}_{x_1,\ldots, x_\nu}
&=& \exp\left[- \hf \mr{Tr} \ln \tG^{-1}
 - \hf \tbe^2 \sum_{i\not=j}^\nu \sigma_i\sigma_j \tG_{x_i,x_j}
  \right]\,,\\[2ex]
w_\sigma &=&  -\hf a^d u_\sigma \exp\left[-\hf \sigma^2 \tbe^2 \tG (a) \right]
\end{eqnarray}
\end{subequations}
with the dimensionless coupling $\tbe^2= a^{-( d-2)} \beta^2$
and the dimensionless propagator 
\begin{equation}
  \tG_{x,y}= \left\{ 
        \begin{array}{lll}
        - c_2 \, \ln \left(\displaystyle {|x-y|\over {\bar a}}\right)& 
        \mbox{ for }& d=2\,,\\[4ex]
        c_d \, {\displaystyle {\bar a}^{d-2}\over \displaystyle (d-2)|x-y|^{d-2}}& 
          \mbox{ for }& d\ge 3\,.
        \end{array} \right.
\end{equation}
Here, $c_d=\Omega^{-1}_d$ where $\Omega_d$ 
denotes the entire solid angle in $d$ dimensions, and ${\bar a}
=\ord{a}$. Its explicit value has been determined for $d=2$ (c.f. \cite{gulacsi}).
For the sake of simplicity we shall take here ${\bar a}=a$. The integration regions 
on the r.h.s. of Equation \eq{zcg}  are discussed in 
App. \ref{renecg}. Eq. \eq{zcg} represents the grand-canonical partition 
function for a gas of static Coulomb charges. The  temperature of this 
ECG is given as $\propto (\beta^2)^{-1}$, and the Fourier amplitudes 
$u_\sigma$ of the periodic self-interaction potential in the GSGM   
play the role of the fugacities of  the various charges 
$\sigma=\pm 1, \pm 2,\ldots$; this mapping 
of the partition function of the GSGM to that of the ECG is exact for 
arbitrary dimensionality.

In this work we shall check our RS-RG results for the ECG
in higher dimensions against the RG-flow equations obtained in \cite{Nienhuis} 
with the same approach for $d=2$. It is a little cumbersome to 
identify the various notations (see Tab.~\ref{tab1} for an explanation).
Indeed, the two-dimensional ECG is an electric Coulomb gas in the sense of 
\cite{Nienhuis}. 
Applying the RG flow equations \cite[Eqs.~(2.22) and (2.23)]{Nienhuis} 
to the two-dimensional ECG with $Z(2)$ symmetry 
($w_\sigma=w_{-\sigma}$),  one obtains: 
\begin{eqnarray}\label{niencg}
&& a{d\tbe^2\over da} 
= -\pi \tbe^4 \sum_{\sigma=1}^\infty \sigma^2 w_\sigma^2,
~~~~
 a{dw_\sigma\over da} =
 \biggl( 2- {\tbe^2\over 4\pi} \sigma^2 \biggr) w_\sigma
   + \pi \sum_{\sigma'\not=0,\sigma} w_{\sigma'}w_{\sigma-\sigma'}
    - \pi  \biggl( {8\pi\over 3} +\sqrt{3}\biggr) w_\sigma 
      \sum_{\sigma'=1}^\infty w_{\sigma'}^2.
\end{eqnarray} 
From the linearized flow equations one recovers the Coleman fixed 
point \cite{col} of the sine-Gordon model at $\tbe^2=8\pi$.
Below, we also compare our RS-RG flow equations derived for the ECG for
arbitrary dimensionality to Eq.~\eq{niencg}.

\begin{table}[htb]
\begin{center}
\begin{minipage}{10cm}
\begin{center}
\begin{center}
\begin{tabular}{|c|c||c|c|}
\hline
&&&\cr
\cite{Nienhuis} & ECG& \cite{Nienhuis} & VG\cr
&&&\cr
\hline
&&&\cr
$1/g$ & $ \tbe^2/ (2\pi) $&$g$& ${2\pi J}$  \cr
&&&\cr
$e$&$\sigma$ &$m $& $q$\cr
&&&\cr
$Y(e,0)$&$w_\sigma$ &$Y(0,m)$ & $A_q$\cr
&&&\cr
\hline
\end{tabular}
\end{center}
\caption{\label{tab1} Identification of the various notations 
for the parameters of the two-dimensional ECG (first and second columns),
as well as for the VG (third and fourth columns) according to 
Eqs. (2.1) and (2.2) in \cite{Nienhuis} and our Eqs. \eq{ZGSGM2}, 
\eq{zcg}, and \eq{zvg}. }
\end{center}
\end{minipage}
\end{center}
\end{table}

\subsection{Gas of topological excitations}\label{gtopexc}

In general, dual transformations \cite{savit} can be found and applied 
to almost any Abelian theory in any number of dimensions. The dual form 
of the original theory represents an intermediate step towards its third
form  when the partition function is expressed in terms of the topological 
excitations of the original degrees of freedom (see \cite{gulacsi}),
the gas of topological excitations. 
   
The $d$-dimensional XY-model is given by the partition function
\begin{equation}\label{XY1}
Z_{XY} = \int \cd{\mb{S}_x} \,
\delta ( \mb{S}_x^{\;2} -1)\,
\exp\left[ - \sum_{<y,z>} (-J) \, \mb{S}_y\cdot \mb{S}_z \right]\,,
\end{equation}
where the classical spin $\mb{S}_x$ is a unit-vector in the two-dimensional
internal space; $\sum_{<y,z>}$ stands for the sum over pairs of nearest
neighbour lattice sites. For  the numbers $d=2$ and 
$d=3$ of dimensions the corresponding gases of topological excitations 
are well-known:
\begin{enumerate}
\item {\em The vortex-gas (VG) for  $d=2$ dimensions.}
The partition function of the XY-model can be rewritten as 
(see~\cite{gulacsi})
\begin{equation}\label{zvg}
  Z_{VG} =  \sum_{\nu=0}^\infty {1\over \nu!}
 \sum_{x_1,\ldots,x_\nu}
 A_{q_1}\cdots A_{q_\nu} 
 \exp\left[- \hf{4\pi^2  J} \sum_{i\not =j}^\nu q_iq_j \tG_{x_i,x_j} \right]\,.
\end{equation}
This describes a system of static vortex charges interacting 
via the two-dimensional Coulomb potential 
\[
\tG_{x_i,x_j} = -{1\over 2\pi} \ln (|x_i-x_j|/a)\,,
\]
where the fugacities of the vortices are defined as
$A_q=  \exp[- \mu q^2 J/2]$; $q=\pm 1,\pm,2,\ldots,$
and $\mu$ can be interpreted as the chemical potential of the 
vortex-antivortex pairs. The VG and the ECG are both Coulomb gases, 
but those of different physical origin. The vortices are the 
topological excitations of the free periodic scalar field, whereas 
the Coulomb charges of the ECG are connected with the periodic 
self-interaction of the scalar field. Using the identification of 
various notations given in Table \ref{tab1} (last two columns),
one may interpret the VG as a magnetic Coulomb gas in the sense of 
\cite{Nienhuis}. Again,
applying the RG flow equations \cite[Eqs.~(2.22) and (2.23)]{Nienhuis} 
for the VG with $Z(2)$ symmetry ($A_q=A_{-q}$), one 
finds
\begin{eqnarray}\label{niendcg0}
  a{d J\over da} = - 4\pi^3 J^2 \sum_{q=1}^\infty q^2 A_q^2
,~~~~
 a{dA_q\over da} = \biggl( 2- {\pi} J q^2 \biggr) A_q
+ \pi \sum_{q' \not=0,q} A_{q'}A_{q-q'}
- \pi \biggl( {8\pi\over 3} +\sqrt{3}\biggr) 
  A_q \sum_{q'=1}^\infty A_{q'}^2.
\end{eqnarray}
According to Table \ref{tab1}
 the RG equations \eq{niendcg0} can be 
rewritten in terms of $\tilde \beta^2 = 1/J$ as
\begin{eqnarray}\label{niendcg}
  a{d \tilde\beta^2  \over da} = 4\pi^3 \sum_{q=1}^\infty q^2 A_q^2,
~~~~
a{dA_q\over da} = \biggl( 2- {\pi\over \tilde\beta^2} q^2 \biggr) A_q
+ \pi \sum_{q' \not=0,q} A_{q'}A_{q-q'}
- \pi \biggl( {8\pi\over 3} +\sqrt{3}\biggr) 
  A_q \sum_{q'=1}^\infty A_{q'}^2.
\end{eqnarray}

\item {\em The vortex-loop gas (VLG) for $d=3$ dimensions.}
For the three-dimensional XY model the dual 
theory has been constructed in the literature \cite{nelson,shenoy} 
and it turned out to be the gas of interacting vortex loops. Also 
the corresponding flow equations have been derived for the parameters 
$K \propto J$   and  $y$ (the fugacity of the 
vortex loops) by RS-RG in the DGA:
\begin{eqnarray}\label{XYdual}
  a{dK\over da} = K-  {4\pi^3\over 3} K^2 y,~~~~
 a{dy\over da}= \biggl( 6- \pi^2 K L \biggr) y,
\end{eqnarray}
where $L \equiv L(a)$ corresponds to the long-range contribution.
One may assume $L$ to approach a constant in the IR limit,
or to be weakly (logarithmically) divergent,
e.g. $L(a) = 1$ in Ref.~\cite{nelson}, $L(a) = \ln(a/a_c ) + 1$ 
in Ref.~\cite{shenoy}.
\end{enumerate}

%
%
\section{RG flow of GSGM}\label{flowGSGM}

\subsection{Differential RG in momentum space}

Below we shall discuss the renormalization of the GSGM 
(Euclidean one-component scalar field theory with periodic 
self-interaction) by means of the differential renormalization group 
approach in momentum space. In this approach, blocking transformations 
\cite{wilson} are realized by successive elimination of the field 
fluctuations according to their decreasing momentum in infinitesimal 
steps. The physical effects of the eliminated modes are encoded in the
dependence of the coupling constants on the scale $k$ above which
the high-frequency modes are effectively eliminated. The elimination 
of the modes above the moving scale $k$ is either complete as in 
Wegner's and Houghton's method (WH-RG) \cite{wh} (sharp IR cut-off)
or partial as in Polchinski's method \cite{polc} (smooth IR cut-off).
Both of these RG methods keep intact the periodicity, 
which is the essential symmetry 
of theory. Nevertheless, choosing the WH-RG method is a 
necessity in our case: as shown below, a spinodal instability occurs 
at some critical momentum scale  $k_c$. Therefore, the elimination of the 
modes above and below this critical scale involves completely different 
physics. Renormalization of the couplings occurs due to small
quantum fluctuations above the critical scale, whereas tree-level
fluctuations of large amplitude dominate below that scale \cite{tree}. 
Thus, the application of the RG methods with smooth cut-off like that 
of  Polchinski's method \cite{polc,comellas,ball} is now unjustified, 
because such methods treat the modes at all scales similarly. 

Above the critical scale $k_c$
the WH-RG method \cite{wh} provides an  integro-differential equation
for the blocked action $S_k[\phi]$ as the functional of the scalar 
field $\phi$. In order to solve this equation, one  has to project it 
to a particular functional subspace. Therefore, one generally assumes 
that the blocked action contains only local interactions, expands it 
in powers of the gradient of the field, and truncates this expansion 
at a given order, for technical reasons. 
The drawback of the WH-RG method combined with 
the gradient expansion is that the truncation of the latter may introduce 
an unphysical high-frequency behaviour of the blocked action, since a 
sharp cut-off is used \cite{morris}. 
Here, we restrict ourselves to the 
leading order of the gradient expansion, the local-potential approximation 
(LPA). In order to get more insight in the RG evolution of the blocked 
action, it may turn out to be useful in the future 
to replace the gradient expansion with some more 
appropriate approximation scheme, like the cluster expansion proposed 
in \cite{kornelsajat}.
In LPA the ansatz for the blocked action can be written as 
\begin{equation}
S_k = \, \int {\mathrm d}^d x \left[  
 \hf \, (\partial_{\mu}\phi)^2 \, + \, V_k (\phi) \right]
\end{equation}
with the blocked potential $V_k (\phi)$ for which the WH equation  
reads as follows \cite{O(N),senben,hh,janosRG,ber}:
\begin{equation}
\label{WH}
k\partial_k V_k(\phi) = - k^d \alpha_d 
\ln \left({ k^2+\partial^2_{\phi} 
V_k(\phi) \over  k^2} \right),
\end{equation}
with $\alpha_d = {1\over2} \Omega_d (2\pi)^{-d}$, where $\Omega_d$
denotes the solid angle in $d$ dimensions. In order to look for 
fixed-point solutions of \eq{WH} or to follow the scaling of an 
arbitrary potential, it is convenient to remove the trivial scaling 
of the dimensionful coupling constants and rewrite Eq.~\eq{WH} as
\begin{equation}
\label{WHdim}
\left(d - \frac{d-2}{2} {\tilde\phi} \partial_{\tilde\phi}  
+ k\partial_k \right) {\tilde V_k({\tilde\phi})} = 
\,- \hbar \, \alpha_d \, \ln\left(1 +\partial^2_{\tilde\phi} 
{\tilde V_k({\tilde\phi})}\right)\,,
\end{equation}
where the dimensionless quantities
${\tilde\phi} = k^{-{d-2\over2}}\phi$ and 
${\tilde V_k({\tilde\phi})}=k^{-d}V_k(\phi)$ have been introduced.

Notice that the argument of the logarithm in Eqs. \eq{WH} and 
\eq{WHdim} must be positive. If the argument vanishes or if it 
changes sign at a 
critical value $k_{\mathrm c}$, given by 
$k_{\mathrm c}^2 = -\partial^2_{\phi} V_{k_{\mathrm c}}(\phi)$,  
the WH equation \eq{WH} loses its validity for $k<k_{\mathrm c}$.
This is a consequence of the spinodal instability. At the scale
$k_{\mathrm c}$ zeros occur  among the eigenvalues of the second
derivative matrix of the blocked action, the restoring force
vanishes for some modes of the fluctuations and the amplitudes of such 
fluctuations can grow to a large finite value and cause a tree-level 
renormalization \cite{tree}. Then, the tree-level blocking relation 
\cite{tree}
\begin{equation}
\label{treeLO}
V_{k-\delta k}(\phi) 
= \min_{\rho} \left[k^2 \rho^2 + {1\over 2} \int_{-1}^{1} 
{\mathrm d} u \hskip 0.2cm V_{k}(\phi + 2\rho \cos(\pi u)) \right],   
\end{equation}
should be used instead of Eq. \eq{WH} for the determination of the 
blocked action $V_{k-\delta k}(\phi) $ at the lower scale $k-\delta k$.

%
%
\subsection{Global aspects of the renormalization of the GSGM}\label{global}

As it is argued in Refs. \cite{per,jpg} for $d=2$ dimensions, the 
WH-RG procedure retains the periodicity of the potential. 
The argumentation holds for any number of dimensions: 
both Eqs.~\eq{WHdim} and \eq{treeLO} which 
describe the infinitesimal blocking step $\delta k$
can be considered in the form 
$V_{k-\delta k}(\phi)= {\cal F}[ V_k(\phi)]$ with some functional 
${\cal F}$. If the r.h.s. of this equation is periodic at the scale 
$k$ due to 
$V_k(\phi)=V_k(\phi +\Delta)$, then its l.h.s., i.e.~the blocked potential 
$V_{k-\delta k}(\phi)$, should also be periodic with the same period 
$\Delta$. This means that (i) the WH-RG in LPA keeps periodicity of the 
blocked potential, and (ii) the period $\Delta$ in the internal space 
does not depend on the scale. Note that these statements are only 
valid for the dimensionful potential as the local functional of the
dimensionful field. 

Since the blocked potential $V_k(\phi)$ tends to the effective 
potential $V_{\mathrm {eff}}(\phi)$ in the IR limit $k\to 0$, the 
effective potential should be periodic, as well. On the other hand, 
the effective potential is convex \cite{conv} over the homogeneous
field configurations. The double requirement of periodicity and 
convexity can only be fulfilled in a trivial manner,  namely if 
$V_{\mathrm {eff}}(\phi)$ is a constant function. In order to 
demonstrate  this general statement, we determine the blocked 
potential for the $4$-dimensional GSGM numerically.

\subsection{RG equations}

Here, we  apply  the flow equations \eq{WHdim} and \eq{treeLO} 
for the GSGM. For the sake of simplicity, we consider periodic 
potentials with Z(2) symmetry, $V_k(\phi) = V_k(-\phi)$. The 
Fourier-expansion of such  potentials is given by
\begin{equation}
\label{pot-Fourier}
{V_k(\phi)} = \sum_{n=1}^{\infty} {u_n(k)} 
\cos\left(n\beta\phi\right),\hskip 1cm
{\tilde V_k(\tilde\phi)} = \sum_{n=1}^{\infty} {\tilde u_n(k)} 
\cos\left(n {\tilde\beta} {\tilde\phi}\right)
\end{equation}
where $\beta=2\pi/\Delta$ is independent of the scale $k$, and the 
tilde indicates the corresponding dimensionless quantities.

Inserting the Fourier-expansion \eq{pot-Fourier} into 
Eq. \eq{WHdim} differentiated 
with respect to $\tilde\phi$, one obtains an equation with terms either 
non-periodic or  periodic in the field variable $\phi$. The non-periodic 
terms occur due to the scale-dependence of ${\tilde\beta}(k)$ and  the 
term $- \frac{d-2}{2} {\tilde\phi} \partial_{\tilde\phi} 
{\tilde V_k(\tilde\phi)}$
in Eq. \eq{WHdim}. The obtained flow equation is satisfied if and only 
if the non-periodic and the periodic parts are identical on both sides 
separately. The non-periodic parts yield:
\begin{equation}
\label{vanish}
\sum_n {\tilde u_n(k)} \, n \, \left( k \partial_k {\tilde\beta}(k)
\, - \, \frac{d-2}{2} \, {\tilde\beta}(k) \right) \, 
{\tilde\phi} \, \sin (n \, {\tilde\beta} \, {\tilde\phi}) = 0.
\end{equation}
Using this relation and the equation for the periodic terms, one gets the 
evolution equations for  arbitrary number $d$ of dimensions as given 
previously in Ref. \cite{per}:
\begin{equation}\label{generalv}
(d + k\partial_k) v_n(k) = 
\alpha {\tilde\beta}^{2}(k) n^2 v_n(k) - 
{1\over 2} {\tilde\beta}^{2}(k)  \sum_{p=1}^{N} A_{n, p}(k) 
(d + k\partial_k) v_p(k),
\end{equation}
\begin{equation}\label{wtbe}
 k \partial_k {\tilde\beta}^{2}(k) = 
({d-2}) \, {\tilde\beta}^{2}(k),
\end{equation}
where $v_n(k) = n {\tilde u}_n(k)$, and
$ A_{n,p}(k)= |n-p|v_{n-p}- (n+p) v_{n+p}$ [note that 
in our previous definitions~\cite{per}
the analog of the first relation is $v_n(k) = n \beta{\tilde u}_n(k)$].
Eq. \eq{wtbe}  follows from Eq. \eq{vanish} and has the solution
\begin{equation}
\label{scalebeta}
{\tilde\beta}^2(k) = \, k^{d-2} \, \beta^2.
\end{equation}
This expresses that the periodicity of the blocked potential
is kept, whereas  the dimensionful period $\Delta ={2\pi\over \beta}$
remains independent of the scale.

In order to solve the flow equations \eq{generalv}, 
one has to express the derivatives $k\partial_k \tu_n(k)$ occurring in
Eq.  \eq{generalv} explicitly and truncate the Fourier-expansion of the
potential at some mode $n=N$. This can be achieved by the following steps:
\begin{enumerate}
\item
Rewrite Eq. \eq{generalv} in terms of $\tu_n={1\over n} v_n$:
\begin{eqnarray}\label{gentu1}
&&\sum_{p=1}^\infty \biggl( \delta_{n,p} + \hf \tbe^2 \tA_{n,p}
    \biggr)
 k\partial_k \tu_p(k) = -d\tu_n(k)
+\alpha_d {\tilde\beta}^{2}(k) n^2 \tu_n(k) 
- {d\over 2} {\tilde\beta}^{2}(k)  \sum_{p=1}^{\infty} \tA_{n, p}(k)  
\tu_p(k)\,,
\end{eqnarray}
with $\tA_{n, p}(k)={p\over n}[ (n-p)^2 \tu_{n-p}  - (n+p)^2 \tu_{n+p} ]$.
\item Invert the matrix on the l.h.s. of Eq. \eq{gentu1} 
numerically in the subspace of the modes with $n,p\le N$.
\end{enumerate}
For later comparison with the flow equations for the $d$-dimensional ECG,
it is useful to perform the inversion of the matrix $1+\hf\tbe^2 \tA$
keeping the terms up to the order  $\ord{\tA^2}$, i.e. those up to
$\ord{\tu^3}$ :
\begin{eqnarray}\label{gentu2}
  k\partial_k \tu_n (k) =  -[d 
-\alpha_d {\tilde\beta}^{2}(k) n^2] \tu_n(k)
-{ \alpha_d\over 2} \tbe^4 \sum_{p=1}^\infty \tA_{n,p} p^2 \tu_p
+{ \alpha_d\over 4} \tbe^6 \sum_{p,r=1}^\infty \tA_{n,r}\tA_{r,p} 
p^2 \tu_p +\ord{\tu^4}.
\end{eqnarray} 

In the region of the  spinodal instability for $k<k_c$, 
the RG flow should be determined with the help of
 the tree-level blocking relation 
\eq{treeLO} which can be rewritten for the dimensionful periodic potential 
as
\begin{equation} 
\label{tree-periodic}
V_{k-\delta k}(\phi) = \min_{\rho} \left[k^2 \rho^2 + \sum_{n=0}^{\infty} 
u_n(k) \cos(n \beta\phi) J_{0}(2 n \beta \rho)\right],   
\end{equation}
where $J_{0}$ stands for a Bessel function, 
$\beta=$constant and $\phi$ is a homogeneous field configuration. It 
is worthwhile to note that Eq. \eq{tree-periodic} is just the same for 
any number $d$ of dimensions. Therefore, it can always be solved in  the 
same manner as discussed in \cite{per} for the case with $d=2$. The 
position $\rho= \rho_k(\phi)$ of the minimum should be determined 
separately for 
each constant field configuration $\phi$ at the given scale $k$. Then, the 
r.h.s. of Eq. \eq{tree-periodic} at the minimum as a function of $\phi$
should be reexpanded in Fourier-modes, in order to determine the coupling 
constants $u_n(k-\delta k)$ of the blocked potential at the lower scale 
$k-\delta k$. The pure tree-level scale-dependence \eq{scalebeta} should 
hold
in both regions $k>k_{\mathrm c}$ and $k<k_{\mathrm c}$ alike.

\subsection{UV scaling}

The trivial Gaussian fixed point ${\tilde V^{*} = 0}$ is the only one of 
the Wegner--Houghton RG equation \eq{WHdim} in the subspace of periodic 
potentials. 
This is because the 
dimensionless coupling  $\tbe(k)$ can only be independent of the scale 
for a periodic potential that vanishes identically,
as Eq. \eq{scalebeta} should hold for a constant $\beta$.

\begin{figure}
\begin{center}
\begin{minipage}{11cm}
\begin{center}
\epsfig{file=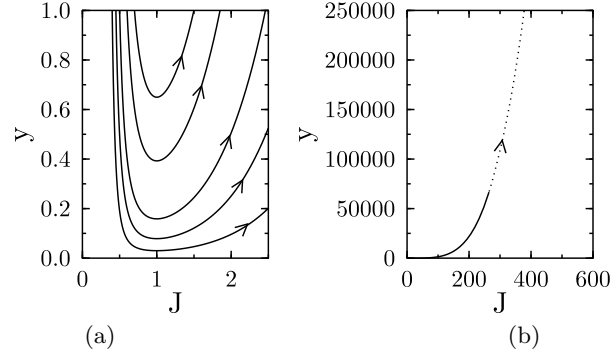, width=8cm}
\hspace{5cm} (a) \hspace{5cm} (b) \hfill\\
 \caption{
 RG flow for the three-dimensional GSGM, obtained by 
 solving  the WH-RG equations (a) Eqs. \eq{wtbe},  \eq{lin-general},
  and (b) Eqs. \eq{wtbe}, \eq{gentu1} outside the
 region of spinodal instability (solid line) and Eqs. \eq{wtbe},
 \eq{tree-periodic} inside the region of the spinodal instability (dotted line) by 
 a fourth-order Runge-Kutta method. 
 The RG trajectories are projected onto the 
 subspace of the dimensionless coupling constants 
 $y = \tilde u_{1}^{2}(k)$ and $J = {1 / \tilde\beta^2(k)}$.
 The lines  represent the RG trajectories for various initial 
 conditions, and the arrows indicate the direction of the RG flow.}
 \label{linsg}
\end{center}
\end{minipage}
\end{center}
\end{figure}

In order to determine the scaling operators and the critical exponents 
of the periodic field theory around the Gaussian fixed point for 
$d\ge 3$ dimensions, one can use the quasi-linear version of the flow 
equation \eq{gentu2}:
\begin{eqnarray}
\label{lin-general}
(d + k\partial_k) \, {\tilde u_n(k)} = \, \alpha_d \, 
{\tilde\beta^{2}} \,  n^2 \, {\tilde u_n(k)} .
\end{eqnarray}
Together with Eq. \eq{scalebeta} this has the analytic solution 
\begin{equation}
\label{solution}
{\tilde u}_n(k) = {\tilde u}_{n}(\Lambda)  
\left({k}\over{\Lambda}\right)^{-d} 
\exp \biggl\{\frac{\alpha_d \tbe^2(\Lambda) n^2}{d-2} 
\biggl[ \biggl( {k\over \Lambda}\biggr)^{d-2} - 1\biggr] \biggr\}
\end{equation}
where $\tbe(\Lambda)$ and ${\tilde u}_{n}(\Lambda)$ are the bare
values of the couplings. Let us consider the solution \eq{solution} 
in the neighbourhood of the UV cut-off $\Lambda$. Introducing a
small positive parameter $\epsilon= 1-{k\over \Lambda}>0$,
the term $({k\over \Lambda})^{d-2}$ can be expanded 
around $\epsilon = 0$:
$(1-\epsilon)^{d-2} = 1 - (d-2)\epsilon + \ord{ \epsilon^2 }$.
Inserting this into the exponent on the r.h.s. of Eq. \eq{solution} 
and using $\ln (1 - \epsilon) = -\epsilon + \ord{ \epsilon^2 }$, 
one finds 
\begin{equation}
\label{exp-scal}
\exp \left(- \alpha_d \, \tbe^2(\Lambda) n^2 \epsilon  + 
\ord{ \epsilon^2 } \right) 
= \exp \left( \alpha_d \, \tbe^2(\Lambda) n^2 \ln (1-\epsilon) + 
\ord{ \epsilon^2 } \right)  \approx
(1-\epsilon)^{\alpha_d \, \tbe^2(\Lambda) n^2 }.
\end{equation}
Then, the solution \eq{solution} can be recast in the form
\begin{equation}
\label{solution2}
{\tilde u}_n(k) = {\tilde u}_{n}(\Lambda)  
\left({k}\over{\Lambda}\right)^{-d} 
\left({k}\over{\Lambda}\right)^{\alpha_d \, \tbe^2(\Lambda ) n^2  + 
\ord{ \epsilon^2 }}. 
\end{equation}
One may be tempted to conclude that
the couplings ${\tilde u}_n(k)$ are relevant,
marginal, and irrelevant depending on  
$\tbe^2(\Lambda) n^2-{d\over \alpha_d}<0$, 
$=0$, and $>0$, respectively. This distinction is, however, made 
according to the {\em bare} value of the coupling $\tbe^2(\Lambda)$. 
The 
latter goes to infinity in the UV regime (pushing the UV cut-off 
$\Lambda$ to infinity), so that in fact {\em all} couplings 
${\tilde u}_n (k)$ are 
irrelevant in the UV regime. As the scale $k$ is 
decreased, a progressively larger number of couplings with 
increasing index $n$ 
become relevant. For example,
as Fig. \ref{linsg}~(a) illustrates, the coupling, 
${\tilde u}_1(k)$ which is irrelevant in the UV regime, becomes 
relevant
along the RG trajectories. In the cross-over region the 
UV scaling laws lose their validity due to the non-linear terms of 
higher order. Including the latter and following the RG trajectories
through the critical scale $k_{\mathrm c}$ down to $k\to 0$,
one does not find any qualitative change in the behaviour of the
RG trajectories. This is illustrated in  Fig.\ref{linsg}~(b) 
using the example of a single RG trajectory (see Sect. \ref{effep}
for the details of the calculation).

The UV scaling behaviour found for $d\ge 3$ is completely different
from that obtained for $d=2$ in Ref.~\cite{per} [see Eq. (17) {\em ibid.}].
For $d=2$, a strong and a weak coupling phases with different scaling 
laws have been detected in the UV regime depending on the actual value 
of the scale-independent parameter $\beta$. For $d\ge 3$ the same UV 
scaling laws have been found for arbitrary values of $\beta$. Neither the UV 
scaling laws, nor the full RG-flow diagram  indicate the existence of
different phases in the lowest order of the gradient expansion.
The disappearence of the topological phase transition for 
$d\ge 3$ of dimensions is of course related to the
fact that long-range order is no longer prohibited
by the Mermin-Wagner theorem in higher dimensions, and to the 
interplay of the long-range order with the topological 
excitations.

\subsection{Effective potential}\label{effep}

We now determine numerically
the effective potential $V_{\mr{eff}}(\phi)$
for the 4-dimensional GSGM as the limit $k\to0$ of the blocked 
potential 
$V_k(\phi)$. For this purpose the flow equations \eq{wtbe} 
and \eq{gentu1} have to be solved numerically for $k>k_c$ above 
the scale $k_c$ 
of the spinodal instability, and the minimization steps according to 
Eq. \eq{tree-periodic} have to be carried out for $k<k_c$ in the region 
of the
spinodal instability. The Fourier-expansion of the periodic potential 
has 
to be truncated at the term with $n=N$, $N$ fixed.
Monitoring numerically the coupling constants $\tu_n(k)$  
($1 \leq n \leq N$) of the periodic field theory, we compare the 
solution 
with the results obtained by solving the 
WH-RG equation \eq{WH} in LPA for the appropriately chosen polynomial 
potential 
\begin{eqnarray}
\label{polynomial}
V_k(\phi)=\sum_{m=1}^N {1\over (2m)!} g_{2 m}(k) \, \, \phi^{2 m} .
\end{eqnarray}
The polynomial potential was ``matched'' to the periodic one
at the UV scale by equating the bare couplings $g_{2 m}(\Lambda)$
to the corresponding coefficients in the Taylor expansion of the bare 
periodic potential at $\phi=0$:
\begin{eqnarray}
V_{\Lambda}(\phi) = \sum_{n=1}^N u_n(\Lambda) \, 
\cos (n \beta \phi) 
=\sum_{m=1}^N \frac{1}{(2 m)!} \left[(-1)^m \beta^{2m} 
\sum_{n=1}^N u_n(\Lambda) n^{2m} \right] \phi^{2m} \equiv 
\sum_{m=1}^N {1\over (2 m)!} g_{2 m}(\Lambda) \, \, \phi^{2 m}.
\end{eqnarray}
In order to avoid the spinodal instability for the polynomial case,
the initial conditions for the polynomial potential were chosen 
specifically as $g_2(k=\Lambda)= 0.0001$, $g_4(k=\Lambda)=0.1$ and 
$g_m(k=\Lambda)=0$, if $m>4$. In this manner for any given values of 
$\beta^2$, and $N$ we get different sets of initial values 
$\tu_n(\Lambda)$ 
for the Fourier-amplitudes $\tu_n(k)$. For the comparison to the 
corresponding
polynomial potential, the blocked  periodic potential 
is again decomposed into a Taylor-series at each scale $k$.
In this way, as a side-effect, one can investigate the influence of the 
violation of  
periodicity due to the Taylor-expansion on the RG flow of the periodic 
potential. 

Equations~\eq{wtbe} and \eq{gentu1} are solved numerically, starting 
from the UV cut-off $\Lambda=1$ by using a fourth-order Runge-Kutta 
method with the step size $\delta k = 10^{-p} k$ with $p=4$. There 
were no appreciable changes in the numerical results by increasing 
$p$ further. The numerics have been tested by reproducing the results of 
\cite{lowd} for the polynomial case. 
Below the critical scale $k_{\mr{c}}$ where the 
spinodal instability occurs for the periodic case, the RG flow was 
determined by the help of the tree-level blocking relation 
\eq{tree-periodic}. As mentioned above, this equation has  the same 
form for any number $d$ of dimensions.
Therefore, the numerical algorithm used and the function  $\rho_k(\phi)$
providing the minimum are the same as those  for $d=2$ in our previous 
work \cite{per}. Generally, for $N\ge \ord{10}$ Fourier modes
a good convergence has been observed at all scales.
For $d=4$ the RG flow projected onto the plane $(J,y)$  
is qualitatively the same as that for $d=3$  in Fig. \ref{linsg}.
This flow diagram shows that no different phases occur in our 
approximation,
but the UV-irrelevant, dimensionless parameter $y$
becomes relevant in the IR regime.

\begin{figure}
\begin{center}
\begin{minipage}{11cm}
\begin{center}
 \epsfig{file=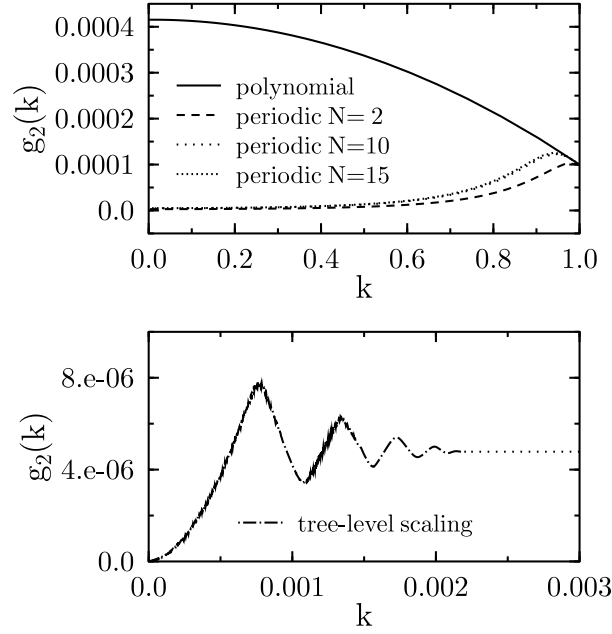, width=8cm}
 \caption{
 RG-evolution of the dimensionful coupling constant $g_2(k)$ versus 
 the running momentum cut-off $k$ (from $k=1 \ldots 0$) in $d=4$ 
 [as a starting point for the RG flow, we fix $\tbe^2(\Lambda=1)$ 
 to the value $128\pi^2$]. Various truncations $N$ of the
 Fourier-expansion of the periodic potential have been employed. 
 The upper and lower plots show the scale-dependence above and 
 below the scale $k_{\mr{c}}$ of the spinodal instability which 
 occurs at $k_{\mr{c}}= 0.0022$  for $N=10$. The solution for the 
 corresponding polynomial case is also indicated 
 (solid line) \cite{tree,lowd}.}
 \label{g2}
\end{center}
\end{minipage}
\end{center}
\end{figure}

\begin{figure}
\begin{center}
\begin{minipage}{11cm}
\begin{center}
 \epsfig{file=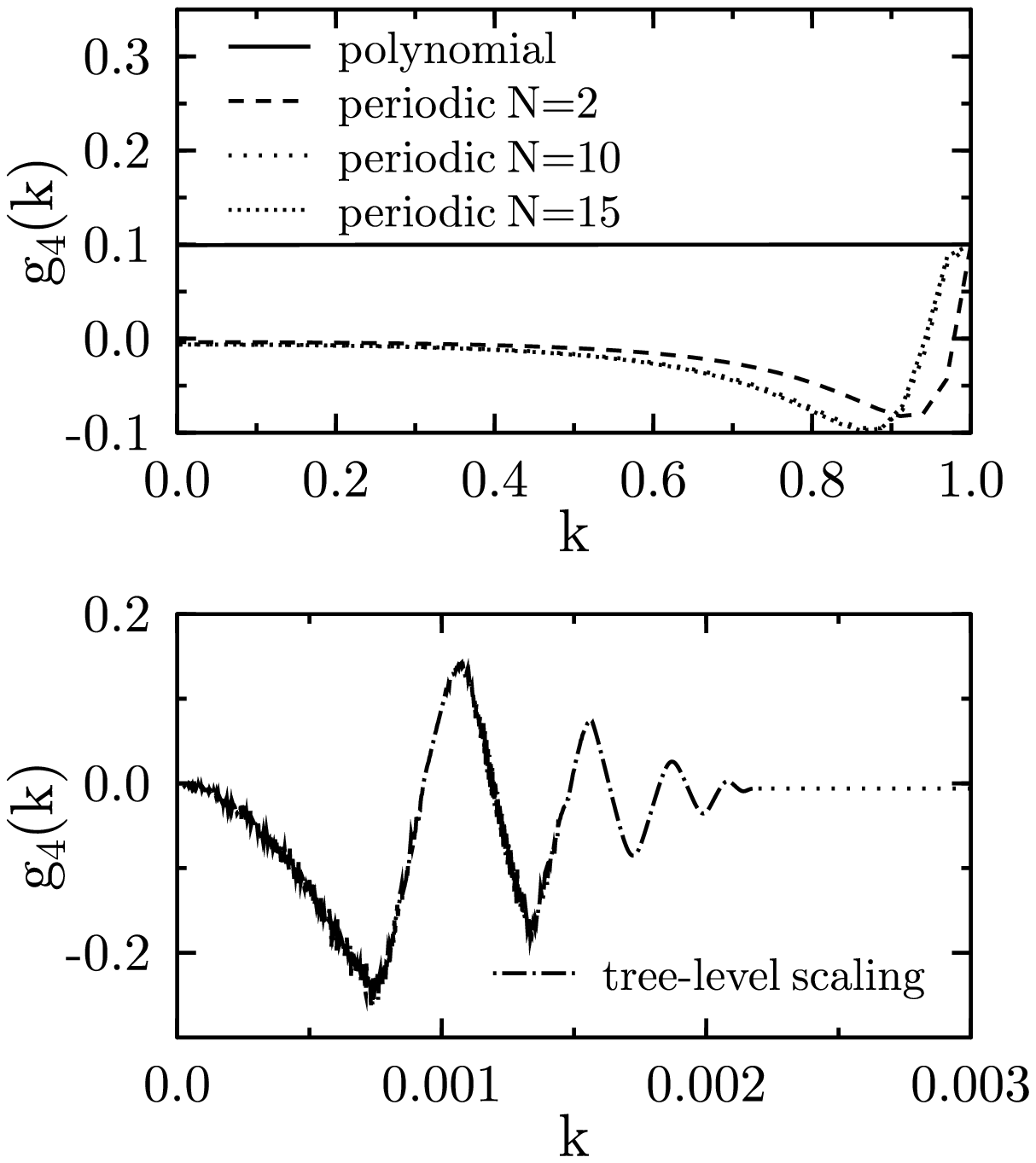, width=8cm}
 \caption{
 The same as Fig. \ref{g2} for the coupling constant $g_4(k)$.}
 \label{g4}
\end{center}
\end{minipage}
\end{center}
\end{figure}

\begin{figure}
\begin{center}
\begin{minipage}{11cm}
\begin{center}
 \epsfig{file=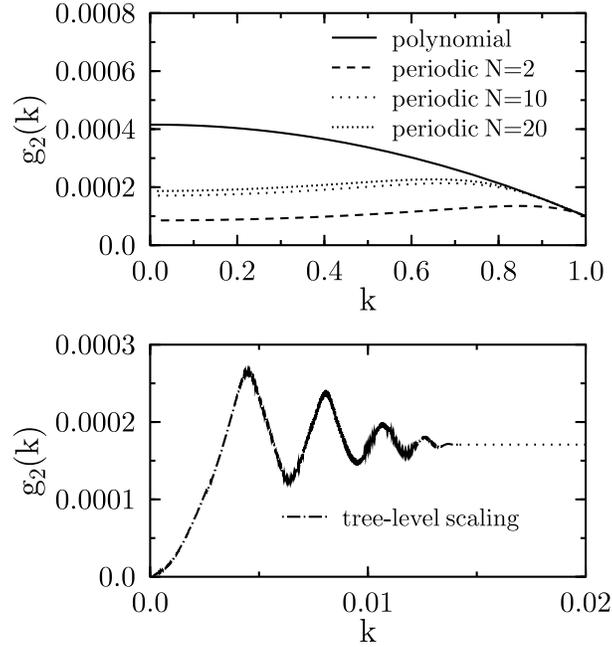, width=8cm}
 \caption{
 The dimensionful coupling constant $g_2(k)$ is plotted versus 
 the running momentum cut-off $k$ (from $k=1$ to $0$) for $d=4$ 
 dimensions, with $\tbe^2(\Lambda=1)= 32\pi^2$. Various truncations 
 $N$ of the Fourier-expansion of the periodic potential are employed. 
 The upper and lower plots show, respectively the scale-dependence 
 above and below the scale $k_{\mr{c}}$ of the spinodal instability 
 which occurs at $k_{\mr{c}}=0.01413$ for $N=10$. The solution for 
 the corresponding polynomial case is also indicated (solid line) 
 \cite{tree,lowd}.}
 \label{g2t2}
\end{center}
\end{minipage}
\end{center}
\end{figure}

\begin{figure}
\begin{center}
\begin{minipage}{11cm}
\begin{center}
 \epsfig{file=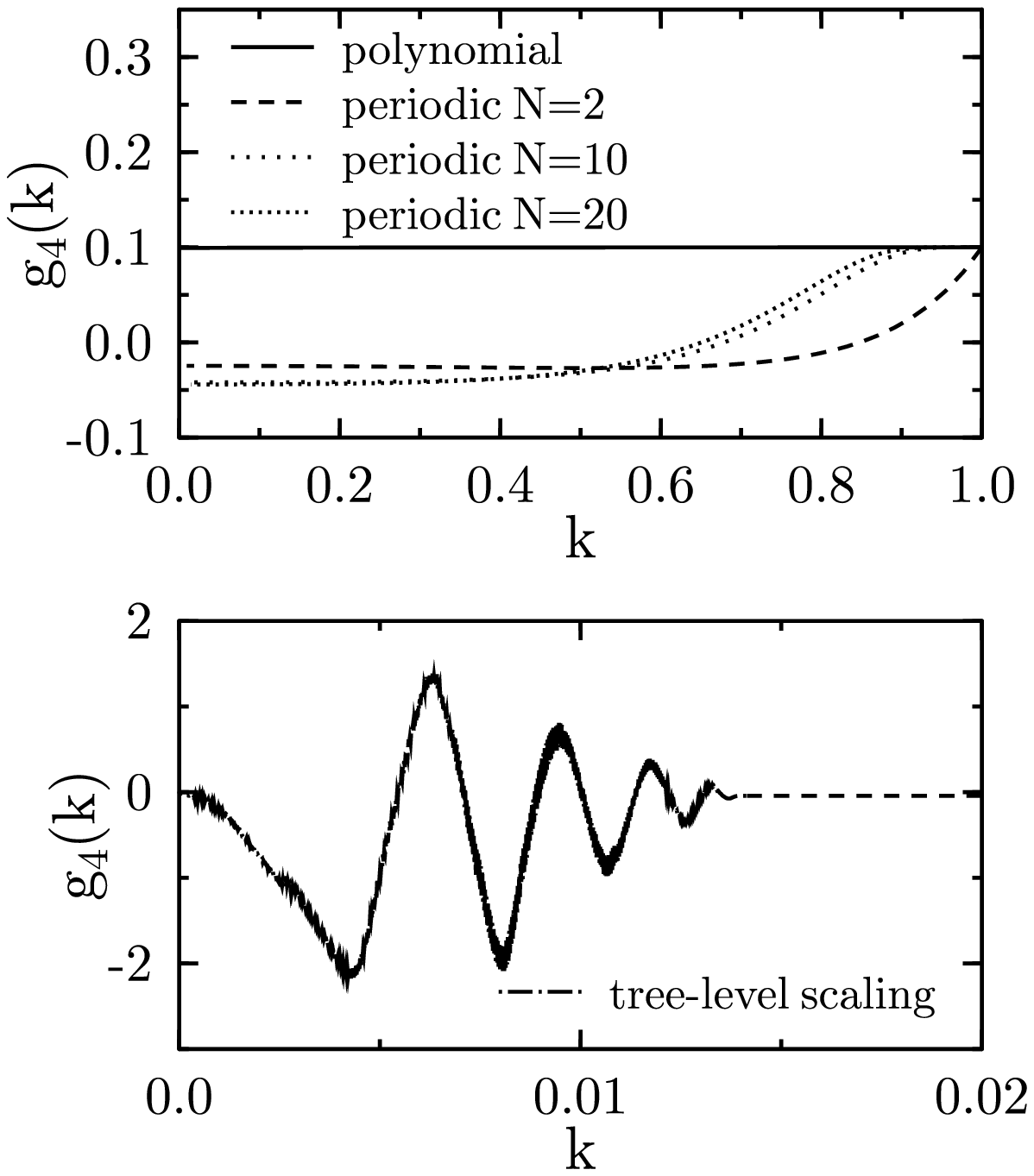, width=8cm}
 \caption{
 The same as Fig. \ref{g2t2} for the coupling constant $g_4(k)$.}
 \label{g4t2}
\end{center}
\end{minipage}
\end{center}
\end{figure}

The upper plots in Figs. \ref{g2} and  \ref{g4}, and 
those in  Figs. \ref{g2t2} and  \ref{g4t2} show the scaling of the 
dimensionful coupling constants $g_2(k)$ and  $g_4(k)$
above the scale of the spinodal instability for various  
truncations $N$ of the Fourier expansion of the potential and for the 
bare values $\tbe^2(\Lambda)=2\tbe_0^2= 128\pi^2$ and $\tbe^2(\Lambda)=
\hf\tbe_0^2=32\pi^2$,
respectively. [According to Eq.~\eq{solution2}, $\tu_1(k)$  is 
independent
of the scale $k$ for $\tbe^2 (\Lambda)=\tbe^2_0 =  64 \pi^2$.] We see 
the good 
convergence of the numerical results with increasing  number $N$ of 
the included Fourier-modes. There is no significant 
difference in the extreme UV regime between the solutions for the 
periodic and the corresponding polynomial cases. Below this 
narrow UV ``window'', the scaling laws for the coupling constants of the
periodic potential deviate remarkably from those for the corresponding
polynomial potential. Namely, the coupling constants for the periodic 
case start to decrease in magnitude after
passing through a cross-over region.

The lower plots  in Figs. \ref{g2}---\ref{g4t2} show the scaling of the 
coupling constants $g_2(k)$ and $g_4(k)$ in the region of the spinodal 
instability for various values of the bare coupling $\tbe^2(\Lambda)$. 
The corresponding curves in the lower plots have not been artificially 
smoothened out in order to demonstrate the relatively small inaccuracy 
of the numerical calculation.
In this deep IR regime the dimensionful coupling constants tend to zero 
for the limit $k\to 0$ in an oscillatory manner. Increasing the number 
$N$ of the encountered Fourier-modes of the blocked potential does not 
alter this general behaviour. This behaviour is qualitatively different 
from that for the polynomial case when all dimensionful coupling constants 
(with even indices) tend to a non-vanishing constant for 
$k\to 0$ \cite{tree,lowd}.

\begin{figure} 
\begin{center}
\begin{minipage}{11cm}
\begin{center}
 \vspace{-12cm}
 \centerline{\epsfig{file=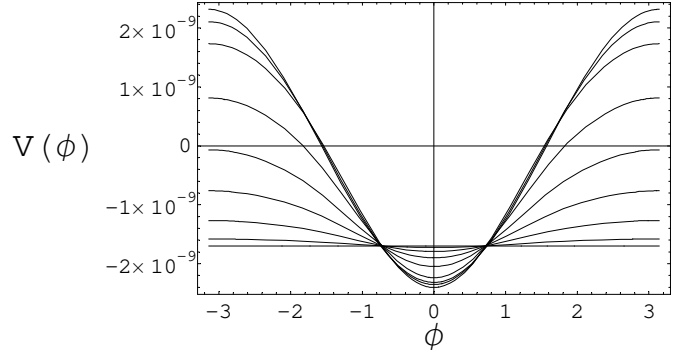, width=22cm}}
 \vspace{-12.5cm}
 \caption{
 The flattening of the dimensionful periodic potential $V_k(\phi)$
 is plotted inside the  region of spinodal instability, $k<k_{\mr{c}}$.
 The various curves with decreasing curvature at $\phi=0$ correspond 
 to different values of the scale decreasing from $k=k_{\mr{c}}$ in
 steps of $\delta k= k_{\mathrm c}/10$.}
 \label{pot}
\end{center}
\end{minipage}
\end{center}
\end{figure}

As illustrated in Fig. \ref{pot}, 
the blocked periodic potential flattens out to a constant
in the limit $k\to 0$. This behaviour
is in accordance with the global features of the RG flow discussed in 
Sect. \ref{global}. In this case the information on the IR physics of 
the system is encoded in the manner in which the blocked potential 
tends to a 
constant value. Because the spinodal instability always occurs, the 
minima 
of the blocked potential are filled due to the large-amplitude 
fluctuations 
of the field.

%
%
\section{RG flow of CG}\label{flowCG}

\subsection{RS-RG for CG}

In this section we apply the RS-RG method in the dilute
gas approximation (DGA) to the renormalization of the
$d$-dimensional ECG. The same approach  was applied
to the two-dimensional generalized Coulomb gas many years ago in
\cite{Nienhuis}. Here, we generalize it for the the partition 
function \eq{zcg}
with in $d\ge 2$ dimensions while following as close as possible 
the derivation in 
\cite{Nienhuis} for $d=2$.  The details of the renormalization 
procedure are described in App.~\ref{renecg}. The RS-RG in DGA 
provides the 
following flow equations for the parameters of the $d$-dimensional 
ECG:
\begin{eqnarray}\label{rswcg}
a {d\tbe^2\over da}  &=&
 - (d-2)\tbe^2 
 + \tbe^4 { \Omega_d^2\over  d}
  \biggl\{ \begin{array}{l}
 - c_2 \cr
  \hf c_d  \end{array}\biggr\}
  <\sigma^2> ,
\nonumber\\
a{dw_\sigma\over da} &=&
 w_\sigma d 
  - \hf c_2 \, \delta_{d,2}\,\tbe^2 \, w_\sigma \, \sigma^2 
+ \hf \Omega_d 
  \sum_{\sigma'\not=\sigma} w_{\sigma'}w_{\sigma-\sigma'}
 \exp\left[ -\hf \tbe^2 \sigma'(\sigma-\sigma') \tG(a)\right]
\nonumber\\
&&
 - w_\sigma {\Omega_d^2 \over d}
  \sum_{\sigma'\not=0} w_{\sigma'} w_{-\sigma'} 
 \exp\left[ \hf \tbe^2 {\sigma'}^{\;2} \tG(a)\right]
 - {d-2\over 4 d}\tbe^4  w_\sigma \Omega_d^2
    \sigma^2   \tG^2(a) <{\sigma'}^{\;2}> \,,
\end{eqnarray}
where
\begin{equation}
<\sigma^2>=  \sum_{\sigma\not=0} \sigma^2 w_\sigma w_{-\sigma}
\exp\left[ \hf \tbe^2 \sigma^2 \tG(a)\right]\,,
\end{equation}
and $\tG(a)=(1-\delta_{d,2}) [(d-2)\Omega_d]^{-1}$.
Note that the first and second rows of the curly bracket on the r.h.s. of the 
first of Eqs. \eq{rswcg} correspond to the cases $d=2$ and $d>2$, respectively.

The well-known flow equations for $d=2$ 
\cite{Nienhuis} are recovered by  these general formulae. For $d=2$ 
and  $Z(2)$ symmetry, $w_{-\sigma}=w_\sigma$, one gets $\tG(a)=0$
and $<\sigma^2>=2 \sum_{\sigma=1}^\infty \sigma^2w_\sigma^2$, and
the RG equations take the form:
\begin{eqnarray}
 a {d\tbe^2\over da} =
 - 2\pi \tbe^4 \sum_{\sigma=1}^\infty \sigma^2w_\sigma^2,
~~~~
 a{dw_\sigma\over da} = \biggl( 2- {\tbe^2\over 4\pi} 
\sigma^2\biggr)w_\sigma + \pi \sum_{\sigma'\not=0,\sigma} 
w_{\sigma'}w_{\sigma-\sigma'}
- 4\pi^2 w_\sigma \sum_{\sigma'=1}^\infty w_{\sigma'}^2.
\end{eqnarray}
Up to numerical prefactors, these equations are  the same as 
Eq.~\eq{niencg} in Sect. \ref{gtopexc}, which represent
the well-known RG equations for the 
two-dimensional electric Coulomb gas \cite{Nienhuis}. The extra factor 
2 on the r.h.s. of the first equation appears due to the contributions 
of both charges $\sigma=\pm 1$ to the value of $<\sigma^2>$.
The other, slightly different factors  depend on the different inaccurate
determinations of the overlapping regions of integrations for momenta, when 
the contributions of the annihilation and fusion of charges are calculated
(e.g. see the note after Eq. \eq{overlapping}).

%
%
\section{Comparison of RG flow for related models}\label{duality}

\subsection{RG flow for GSGM and  ECG}

The flow equations for the $d$-dimensional ECG can be compared to 
those for the GSGM. Such a comparison is meaningful only for the UV 
regime due to different renormalization schemes used in the 
two cases. Furthermore, the RS-RG method has the drawback that is not 
adequate to detect the spinodal instability. This is because the RS-RG 
applies a sharp cut-off in real space which corresponds to a smooth 
cut-off in momentum space. Therefore, there is no way to treat modes 
above and below the critical cut-off $k_c$ differently and to detect 
the spinodal instability in the RS-RG approach. 

\begin{figure}
\begin{center}
\begin{minipage}{11cm}
\begin{center}
 \epsfig{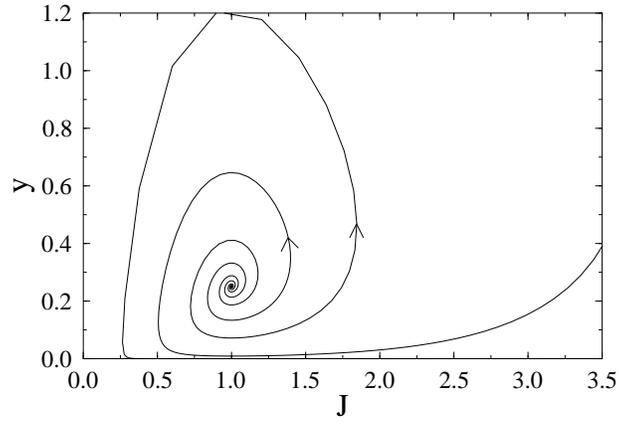}
 \caption{
 RG flow for the ECG  in the plane $(J,y)$, calculated
 without the terms of $\ord{y^2}$ on the r.h.s. of Eq. \eq{DCGKyB}.}
 \label{rscg1wout}
\end{center}
\end{minipage}
\end{center}
\end{figure}

In order to compare the RG flow of the ECG with that of the GSGM, we 
rewrite Eqs. \eq{rswcg} in terms of the dimensionless fugacities
$\tu_\sigma=  a^d u_\sigma  $ by making use of Eq. \eq{wtu}.
We also introduce the momentum scale $k\propto a^{-1}$. Then, the flow 
equations obtained by the RS-RG for the ECG take the form:
\begin{eqnarray}
\label{rscgtbe}
k {d\tbe^2\over dk} &=& + (d-2)\tbe^2 
      - (1-3\delta_{d,2}) {\Omega_d\tbe^4\over 4d}
 \sum_{\sigma=1}^\infty \sigma^2 \tu_\sigma
    \tu_{-\sigma} \exp\left[ -(1-\delta_{d,2})
{\tbe^2\sigma^2\over 2(d-2)\Omega_d} \right]\,, \\
k {d\tu_\sigma\over dk} &=& - \biggl( d- 
    { \tbe^2 \sigma^2    \over 2\Omega_d} \biggr)\tu_\sigma
+ {\Omega_d\over 4} \sum_{\sigma'\not=0,\sigma} \tu_{\sigma'}
\tu_{\sigma-\sigma'}  \exp\left[ -(1-\delta_{d,2}){\tbe^2 \sigma'(\sigma'-\sigma)
     \over 2 (d-2)\Omega_d} \right] 
\nonumber\\[2ex]
\label{rscgtu}
& &  + {\Omega_d^2\over 4d} \tu_\sigma \sum_{\sigma'\not=0} \tu_{\sigma'}
   \tu_{-\sigma'} \exp\left[ - (1-\delta_{d,2}) {\tbe^2 {\sigma'}^{\;2}
 \over 2 (d-2)\Omega_d} \right]\,.
\end{eqnarray}
These equations should be compared with Eqs. \eq{wtbe} and \eq{gentu2} 
for the GSGM. As to the tree-level scaling, i.e. the linearized flow 
equations, the WH-RG LPA and RS-RG DGA give identical results. The 
non-linear term of the order $\ord{\tu}$ on the r.h.s. of Eq. \eq{gentu2}
is identical with the corresponding term on the r.h.s. of Eq. \eq{rscgtu} 
for 
$d=2$ and has a slightly different numerical prefactor for $d\ge 3$. 
The two methods give completely different results concerning the 
non-linear terms of the order higher than $\ord{\tu}$. This happens
because the gradient expansion and the dilute gas approximation 
cannot be  related directly. For comparison, the RG flow may be
calculated for both models in $d=3$, and the RG trajectories 
may be projected onto the plane $(J=1/\tbe^2, y)$ 
where $y= \tu_1^2$ for the GSGM  [Fig. \ref{linsg}(b)]
and $y =(2\pi)^{-2} \tu_1^2 \exp[-\tbe^2/(8\pi)]$ for the ECG 
[Fig. \ref{rscg1wout}]. In the latter case the projected RG trajectories 
are given by
\begin{subequations}
\label{DCGKy}
\begin{eqnarray}
\label{DCGKyA}
 a{dJ\over da} &=& J -{4\pi^3\over 3} y,\\[2ex]
\label{DCGKyB}
 a{dy\over da} &=& \biggl( 6- {1\over 8\pi J}\biggr) y 
- {32 \pi^4 \over 3} y^2 -{\pi^2\over 6 J^2} y^2 .
\end{eqnarray}
\end{subequations}
A significant difference arises due to neglecting the wave-function 
renormalization in the WH-RG approach for the GSGM. Namely, Eq. \eq{wtbe} 
valid  for the scale-dependence of the 'temperature' at any scale 
$k$, does not contain any loop-corrections as 
compared to the corresponding Eq. \eq{rscgtbe} for the ECG. Therefore, 
Eq. \eq{wtbe} does not have any fixed point at finite values of $J$ and $y$ 
(i.e. those of $\tbe^2$ and $\tu_1$) in the flow diagram for the GSGM  
[Fig. \ref{linsg}(b)]. There exists a UV fixed point 
$(J\to 0, y\to \infty)$ and a zero-temperature fixed point 
$(J\to \infty, y\to \infty)$. The phase diagram for the ECG
(Fig. \ref{rscg1wout}) however 
reveals the existence of a repulsive UV  fixed point at finite values 
of $J$ and $y$, when the terms of the order $\ord{y^2}$ are neglected
in Eqs. \eq{DCGKyB}. For small values of the fugacities the non-linear 
terms are negligible and the RG flow looks rather similarly for  both 
models.
The fixed point in Fig. \ref{rscg1wout} maybe artificial because on the 
one hand,
the RS-RG method in the dilute gas approximation is not well suited to 
determine
the RG flow for large values of the fugacity. On the other hand, the 
WH-RG in the
LPA, the validity of which is not restricted to small fugacity values, 
does not 
exhibit any fixed point. Nevertheless the clarification of the role 
of the 
wafefunction renormalization in the WH-RG method is an open question 
and it seems 
not to be possible in the framework of the gradient expansion.

\begin{figure}
\begin{center}
\begin{minipage}{11cm}
\begin{center} 
 \epsfig{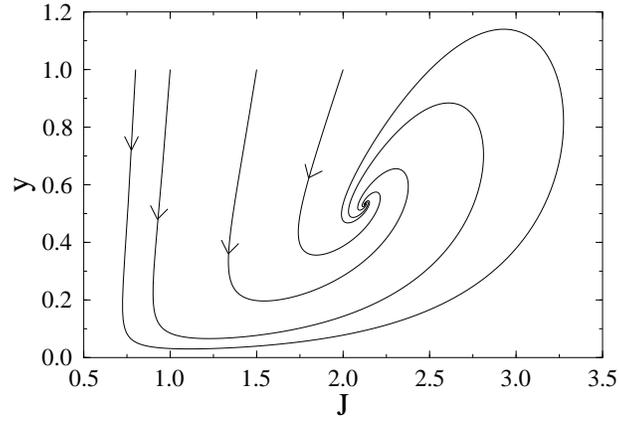}
 \caption{
 RG flow for the ECG  [Eq. \eq{DCGKy}] in the $(J,y)$-plane.}
 \label{rscg1}
\end{center}
\end{minipage}
\end{center}
\end{figure}

The comparison of  the flow diagram in Fig. \ref{rscg1wout} with that 
in Fig. \ref{rscg1} where the terms of the order $\ord{y^2}$
are also included,  reveals the role of the last two 
nonlinear terms on the r.h.s. of Eq. \eq{DCGKyB}. 
Due to these nonlinearities, the fixed point at finite parameter
values turns from a repulsive UV fixed point to an attractive IR one. 
As to the global features of the flow diagram, this indicates that 
one certainly has to go beyond the DGA in order to  clarify the 
essential role of the nonlinearities in the RS renormalization of ECG.

\subsection{RG flow for ECG and VLG}

The RG-flow diagram for the ECG
presented above may be significantly modified
due to non-linear terms of higher order when going beyond the DGA.
Nevertheless,
it can be compared in the UV regime to the RG-flow diagram for the
VLG that was also obtained after certain 
linearizations~\cite{shenoy}.
Such a comparison could in principle give more insight to the question
posed in the literature (see \cite{gulacsi} and the references therein)
on whether the VLG and the three-dimensional XY-model (or the 
three-dimensional
Coulomb gas) really belong to the same class of universality, or not.  
Our strategy is simply to show that there exists a transformation of the 
parameters
$(\tbe^2, \tu_1^2)\mapsto (K,y)$ which transforms the set of 
Eqs. \eq{rscgtbe} and 
\eq{rscgtu} with the charges $\sigma=\pm 1$ for the three-dimensional
 ECG to the equations of the form very close to that of the 
set of Eqs. \eq{XYdual} for the VLG up to the non-linear terms of 
higher order. 
In fact, introducing the parameters
\begin{equation}
   K = \biggl({k_0\over k}\biggr)^2 \tbe^2 
= \biggl({a\over a_0}\biggr)^2 \tbe^2,~~~~
   y = - {1\over 4\pi^2} \tu_1^2 \exp\left[-{\tbe^2\over 8\pi}\right],
\end{equation}
 the flow equations \eq{rscgtbe}, \eq{rscgtu} give 
\begin{equation}\label{dualofdcg}
a{dK\over da} = K - {4\pi^3\over 3} K^2 y  \left({a_0\over a}\right)^2,
~~~~ a{dy\over da}= 
\biggl( 6 - {K \over 8\pi}\left({a_0\over a}\right)^2 \biggr) y 
+\ord{y^2}.
\end{equation}
These equations are very similar to Eq. \eq{XYdual}, although there is
an additional scale dependence in the equation for the flow of the parameter
 $K$. More questionable is the interpretation of $y$ as fugacity of anything 
because it is a negative quantity.
These discrepancies may reflect the dependence on the different
approximations used when the RS-RG has been applied to ECG and VLG, 
respectively.
Therefore, any conclusion on whether the
VLG and the three-dimensional Coulomb gas belong to the same universality 
class,
has to be seen as premature at this stage.

%
%
\section{Summary}\label{summar}

A renormalization-group analysis of the generalized sine-Gordon model 
(GSGM) and
that of the equivalent Coulomb gas (ECG) have been pursued 
for $d\ge 3$ dimensions. The RG equations for the GSGM
and the ECG are derived in two ways:
(i) by means of Wegner's and Houghton's 
(WH-RG) method in the local-potential approximation and (ii) by the 
real-space 
RG (RS-RG) in the dilute-gas approximation. These different approaches 
complement one another: the WH-RG applied to the GSGM is better suited to 
determine the IR behaviour of the system, whereas the RS-RG applied to 
the ECG is able to better represent the non-linearities which influence 
the UV 
scaling. It has been shown that the leading-order terms of the flow 
equations obtained by these methods agree well. Taking as an 
example the $4$-dimensional GSGM, we have demonstrated numerically that 
the 
blocked potential tends to a constant effective potential, independent 
of the field-variable, in the IR limit. This behaviour is in agreement 
with the global requirements on the RG flow that constrain the effective 
potential to be periodic and convex at the same time. Comparing the 
evolution of the periodic blocked potential to that of a polynomial 
potential obtained by a Taylor-expansion of the bare periodic potential, 
we show that a violation of the periodicity would lead to  
a completely incorrect IR scaling.  The flow diagram in Fig.~\ref{linsg} 
shows that only a single phase occurs in our approximation, and that
the UV-irrelevant, dimensionless parameter $y$
becomes relevant in the IR regime.

Because the GSGM, the ECG and the XY-model belong to the same universality
class, the determination of the RG flow of these models for the 
three-dimensional case via two different, complementary methods in
the UV regime enables us to compare the UV scaling with that of a
three-dimensional vortex gas (VLG) \cite{shenoy}. It has been established 
that one may transform the flow equations for the ECG to a
form very close to that of the flow equations for the VLG, 
if only the leading order terms  are kept. Nevertheless, the negative 
fugacity
one needs to introduce and the different explicit scale-dependences
occurring in the two sets of the
flow equations do not enable one to make any definitive conclusion 
regarding the
question whether the VLG and the three-dimensional Coulomb gas belong to 
the same universality class.

%
%
\acknowledgments
This work has been supported by the BMBF, by GSI (Darmstadt)
and by the projects OTKA T032501/00, NATO SA(PST.CLG 975722)5066, 
and M\"OB-DAAD 27 (323-PPP-Ungarn). I. N. is also grateful for 
support by DAAD. Part of the computations have 
been performed in the Supercomputer Lab. of the Faculty of Natural 
Sciences of the Univ. of Debrecen grown out from the donation of the 
Alexander von Humboldt Foundation. I.N. and K.S. are indebted 
to Zs. Gul\'acsi and J. Polonyi for extremely useful discussions.
One of the authors (I.N.) would like to thank K. Vad for the helpful 
discussions on the experimental background.

\newpage

\appendix

\section{RS-RG for the CG}\label{renecg}

Here, we perform a real-space renormalization in the dilute gas 
approximation 
(DGA) of the CG given by the partition function \eq{zcg}.
In order to follow as close as possible the procedure 
employed for $d=2$ (see \cite{gulacsi} and references therein), we 
use continuum integrals 
over the infinitely large volume $D_0$, but cut out spheres $K_j$ of 
radii $a$ at each Coulomb charge $\sigma_j$ centered at $x_j$
(this is different from lattice regularization). Thus, the 
regions of integration for the
positions of the charges are $D_0$, $D_1=D_0-K_1$, $D_2=D_0-K_1-K_2,$
$\ldots$, $D_{\nu-1}=
D_0-K_1-\ldots-K_{\nu-1}$ for the charges $\sigma_1$, $\sigma_2$, $\ldots$,
$\sigma_\nu$, respectively.

The RG flow equations can be found by changing the resolution, i.e. the
smallest separation distance from $a$ to $a+\delta a$  at which two charges
can still be detected separately.
This means that if two charges become closer than the
separation distance $a+\delta a$ they will be seen as a single charge
identical to the sum of the two original ones. The partition function
will evolve due to the following reasons:
\begin{enumerate}
\item the rescaling of the length,
\item the fusion of charges  in a single, non-vanishing one,
\item the annihilation, i.e. fusion of charges in a single, neutral one.
\end{enumerate}
These sub-processes acting on the configurations with the number $\nu$
of charges (described by $Z^{[\nu]}$) give contributions 
$\delta_{rs} Z^{[\nu]}$, $\delta_f Z^{[\nu-1]}$, and
$\delta_a  Z^{[\nu-2]}$  to the configurations with the numbers
$\nu$-, $(\nu-1)$-, and $(\nu-2)$ of charges, respectively, i.e. 
\begin{equation}
  \delta Z= \sum_{\nu=0}^\infty \delta_{rs} Z^{[\nu]}+
 \sum_{\nu=1}^\infty \delta_f Z^{[\nu-1]}+
 \sum_{\nu=2}^\infty\delta_a  Z^{[\nu-2]}.
\end{equation}
The  RG idea is realized by recasting the modified partition sum
$Z+\delta Z$ in the original form observed by $Z$, but with 
modified parameters
$\tbe^2$ and $w_\sigma$. Below we 
determine  all these contributions. 
\begin{enumerate}
\item {\underline {Rescaling.}} 
Rescaling means the change of all integration variables $x_j$ to $x_j'=
x_j (1+\delta a/a)$ and increasing the size of all regions
 $D_j$, $K_j$ correspondingly. Then we rewrite the partition sum as 
an integral over the original variables and regions. 
First of all, the rescaling the volume elements $d^d x$ increases them
by $ (\delta a/a)\,d\, d^d x$; this can be merged into a 
modification of the fugacities,
\begin{equation}
\delta_{rs,1} w_\sigma =  w_\sigma d {\delta a\over a}.
\end{equation}
The change of the propagator should be treated differently for $d=2$ 
and $d\ge 3$.

{\bf The case $d=2$.}
The propagator changes according to
\begin{equation}
  \delta_{rs} \tG(x) =- c_2 {\delta a\over a}
\end{equation}
(independent of $x$). In the interaction energy this yields a
contribution
\begin{equation}
  - \hf \tbe^2\sum_{j,i\not=j} \sigma_i\sigma_j (-c_2)
 {\delta a\over a}.
\end{equation}
However, for neutral configurations we have 
$\sum_{j=1}^\nu\sigma_j=0$, i.e.
$\sum_{i\not=j}\sigma_i=-\sigma_j$, so that one gets the energy 
contribution
\begin{equation}
 -\hf c_2 \tbe^2 \sum_{i=1}^\nu \sigma_i^2   {\delta a\over a}   
\end{equation}
which leads to the following modification of the fugacities:
\begin{equation}
  \delta_{rs,2}  w_\sigma= w_\sigma
\left( \exp\left[  -\hf c_2 \tbe^2  \sigma^2 
                 {\delta a\over a} \right] -1\right)
   =  - \hf c_2  \tbe^2   w_\sigma  \sigma^2  {\delta a\over a}.
\end{equation}

{\bf The case $d\ge 3$.} 
The change of the propagator,
\begin{equation}
  \delta_{rs} \tG(x) =  - (d-2)  \tG (x)  {\delta a\over a},
\end{equation}
is proportional to the propagator itself. Therefore, it
can be encoded in the change of the coupling $\tbe^2$:
\begin{equation}
     \delta_{rs}\tbe^2 = - (d-2)\tbe^2   {\delta a\over a}.
\end{equation}

Thus, we get for any numbers $d\ge 2$ of dimensions
\begin{equation}
    \delta_{rs}\tbe^2 = - (d-2)\tbe^2   {\delta a\over a},
~~~~
  \delta_{rs} w_\sigma=  w_\sigma d {\delta a\over a}
 - \hf c_2 \delta_{d,2}\tbe^2 w_\sigma  \sigma^2  {\delta a\over a}.
\end{equation}

\item {\underline {Fusion.}}

Let us denote the sphere of radius $a+\delta a$ centered at $x_j$ by $K_j'$
and the annulus of inner radius $a$ and thickness $\delta a$ centered at $x_j$
by $\delta_j$. Then the regions of intergration in the rescaled partition
function are $D_0$, $D_1=D_0-K_1'+\delta_1=D_1'+\delta_1$, $D_2=D_0-K_1'-K_2'
+\delta_1+\delta_2$, etc. for $x_1$, $x_2$, etc. respectively. The
contributions of both the fusion and annihilation of the point charges 
$\sigma_p$ and $\sigma_q$ are included in the expression
\begin{eqnarray}
\delta_{f,a} Z &=&  \sum_{\nu =0}^\infty
{1\over  \nu!} 
\biggl[
 \sum_{p=2}^\nu \sum_{q=1}^{p-1}
\int_{D_0}{d^d x_1\over a^d}\cdots  \int_{D_{q-2}'}{d^d x_{q-1}\over a^d}
 \int_{D_{q-1}'}{d^d x_{q}\over a^d} \int_{D_{q}'}{d^d x_{q+1}\over a^d}
\cdots  \int_{D_{p-2}'}{d^d x_{p-1}\over a^d} 
\int_{\delta_{p}}{d^d x_{p}\over a^d}
\nonumber\\
&&
 \int_{D_{p}'}{d^d x_{p+1}\over a^d}
 \cdots 
 \int_{D_{\nu-1}'}{d^d x_\nu\over a^d}
 +{\cal O}\biggl( {\delta a\over a}\biggr)^2 
 \biggr]
 \sum_{\sigma_1,\ldots,\sigma_\nu\not=0}w_{\sigma_1}\cdots w_{\sigma_\nu} 
J^{\sigma_1,\ldots ,\sigma_\nu}_{x_1,\ldots, x_\nu} .
\end{eqnarray}
Hence, we have to calculate integrals of the form
\begin{equation}
   I= \int_{D_0-K_z}{d^d x\over a^d} \int_{\delta_x} {d^d y\over a^d}
     f(x,y) \equiv I_1+I_2 ,
\end{equation}
where the point $z$ satisfies either $|x-z|>2a$ for $I_1$, or $a<|x-z|<2a$
for $I_2$. Therefore, the integral for $y$ in $I_2$ extends only over the 
part ${\bar \delta}_x$ of the annulus $\delta_x$ lying outside of $K_z$:
\begin{equation}
 I_1= \int_{|x-z|>2a} {d^d x\over a^d} \int_{\delta_x} {d^d y\over a^d}
 f(x,y),~~~~
I_2 =  \int_{a<|x-z|<2a}  {d^d x\over a^d} \int_{{\bar \delta}_x}
 {d^d y\over a^d} f(x,y).
\end{equation}
Introducing the center of mass coordinate $t=\hf(x+y)$ and the separation
coordinate $s=x-y$, one gets 
\begin{equation}
 I_1\approx \int_{{3\over 2}a<|t|}{d^d t\over a^d} \Omega_d 
{\delta a\over a}     f(t,s=a),~~~~
 I_2 \approx  \int_{a<|t|<{3\over 2}a}{d^d t\over a^d} \hf  \Omega_d  
{\delta a\over a}     f(t,s=a).
\end{equation}
Then, the integral $I$ can be estimated as
\begin{equation}\label{overlapping}
  I=I_1+I_2\approx \biggl[ \int_{{3\over 2}a<|t|}
+\hf \int_{a<|t|<{3\over 2}a}
\biggr] {d^d t\over a^d} \Omega_d {\delta a\over a}  f(t,s=a)
  \approx 
\int_{D_0 -K_z} \omega_d  {\delta a\over a}     f(t,s=a).
\end{equation}
As to the regions of integration, we make another approximation: in
accordance  with the
relation $K_t(\hf a) \subset K_q\cup K_p \subset K_t ({3\over 2}a)$,
(the arguments indicate the radii of the spheres) we replace $ K_q\cup K_p$
by $K_t$. Due to these approximations terms 
which are polynomials in $a$ may be neglected.
Therefore, our approximations will cease to work as $a$ increases,
 so that our results
are restricted to the UV scaling region. In order to take all terms of the
order $\ord{{\delta a\over a}}$ into account, we perform 
a Taylor-expansion of the function $f(t,s)$ in the variable $s$ before 
the integration over the annulus. Thus, we find
\begin{eqnarray}\label{faz}
\delta_{f,a} Z &=&  \sum_{\nu =0}^\infty
{1\over  \nu!} 
\biggl[
 \sum_{p=2}^\nu \sum_{q=1}^{p-1}
\int_{D_0}{d^d x_1\over a^d}\cdots  \int_{D_{q-2}'}{d^d x_{q-1}\over a^d}
 \int_{D_{q-1}'}{d^d t\over a^d} \int_{D_{q}'}{d^d x_{q+1}\over a^d}
\nonumber\\
&&
\cdots  \int_{D_{p-2}'}{d^d x_{p-1}\over a^d} 
 \int_{D_{p}'}{d^d x_{p+1}\over a^d}
 \cdots 
\cdots \int_{D_{\nu-1}'}{d^d x_\nu\over a^d}
 +{\cal O}\biggl( {\delta a\over a}\biggr)^2 
 \biggr]
\nonumber\\
&&
 \sum_{\sigma_1,\ldots,\sigma_\nu\not=0}w_{\sigma_1}\cdots w_{\sigma_\nu} 
\int_{a<|s|<a+\delta a} {d^d s\over a^d} \left.\biggl(
1 +s_\alpha \partial_\alpha^s +\hf s_\alpha s_\beta \partial_\alpha^s
\partial_\beta^s +\ord{s^3}
\biggr)J^{\sigma_1,\ldots ,\sigma_\nu}_{x_1,\ldots, x_\nu}
\right|_{s=a} .
\end{eqnarray}   
In order to evaluate this, we expand  the exponent first,
\begin{eqnarray}\label{jexp}
J^{\sigma_1,\ldots ,\sigma_\nu}_{x_1,\ldots, x_\nu}\equiv J^{[\nu]}
&=&
 J^{[\nu-2]} \exp \biggl\{  
  - \tbe^2\sum_{j\not=p,q}\sigma_j\sigma_p\tG_{x_j,t+\hf s}
  - \tbe^2\sum_{j\not=p,q}\sigma_j\sigma_q\tG_{x_j,t-\hf s}
  -\hf \tbe^2 \sigma_p\sigma_q \tG(s) \biggr\}
\nonumber\\
& \approx&
 J^{[\nu-2]} \exp \biggl\{  -\hf \tbe^2 \sigma_p\sigma_q \tG(a)\biggr\}
\exp\biggl\{ - \tbe^2\sum_{j\not=p,q}\sigma_j\sigma_p
    \biggl[ 1 +\hf s_\alpha \partial_\alpha^t +{1\over 8}s_\alpha s_\beta
     \partial_\alpha^t\partial_\beta^t \biggr] \tG_{x_j,t}
 \nonumber\\
&&
  - \tbe^2\sum_{j\not=p,q}\sigma_j\sigma_q
 \biggl[ 1 -\hf s_\alpha \partial_\alpha^t +{1\over 8}s_\alpha s_\beta
     \partial_\alpha^t\partial_\beta^t \biggr] \tG_{x_j,t}
\biggr\}.
\end{eqnarray}
Expressions \eq{faz} and \eq{jexp} represent the starting point for the
determination of the contribution of the fusion and annihilation of 
charges to the partition sum. In case of fusion, when the charges 
$\sigma_p$ and $\sigma_q$ fuse into a single, non-vanishing charge 
$\sigma_t=\sigma_p+\sigma_q\not=0$, the only contribution of the 
order $\ord{{\delta a/ a}}$ comes from the zeroth power of $s$ 
in \eq{jexp}. Renumerating the charges in the manner
that  the new charge at $t=x_{\nu-1}$ becomes the $(\nu-1)$-th one,
we can write
\begin{eqnarray}\label{fusin}
\int_{D_{\nu-2}'}{d^d t\over a^d}\int_{a<|s|<a+\delta a} {d^d s\over a^d}
   J^{[\nu]}
&\approx&
   J^{[\nu-2]} e^{ -\hf \tbe^2 \sigma_p\sigma_q \tG(a)}
 \int_{D_{\nu-2}'}{d^d t\over a^d}
    e^{ - \tbe^2 \sum_{j\not=p,q} \sigma_j \sigma_t  \tG_{x_j,t} }
   \Omega_d {\delta a\over a}\nonumber\\
&=&
   J^{[\nu-1]}  e^{ -\hf \tbe^2 \sigma_p\sigma_q \tG(a)}
 \Omega_d {\delta a\over a}.
\end{eqnarray}
The other terms with
$\int_{\delta_p} s^m d^d s\sim \ord{ a^m} {\delta a\over a}$ where
$m=1,2,\ldots$ are neglected. Inserting this in the change of the 
partition function due to fusion,
\begin{eqnarray}
\sum_{\nu=1}^\infty \delta Z^{[\nu-1]} &=& 
\sum_{\nu=1}^\infty {1\over \nu!} \hf\nu(\nu-1)
\int_{D_0}{d^d x_1\over a^d}\cdots  
 \int_{D_{\nu-2}'}{d^d x_{\nu-1}\over a^d}
 \sum_{\sigma_1\not=0}w_{\sigma_1}\cdots 
 \sum_{\sigma_{q-1}\not=0}w_{\sigma_{q-1}}
 \sum_{\sigma_{q+1}\not=0}w_{\sigma_{q+1}}
\cdots
\nonumber\\
&&
  \sum_{\sigma_{p-1}\not=0}w_{\sigma_{p-1}}
 \sum_{\sigma_{p+1}\not=0}w_{\sigma_{p+1}}\cdots
\sum_{\sigma_{\nu}\not=0} w_{\sigma_\nu} 
   \Omega_d {\delta a\over a} 
 \sum_{\sigma_{p}\not=0}w_{\sigma_{p}}\sum_{\sigma_{q}\not=0}w_{\sigma_{q}}
  [ 1- \delta_{\sigma_p,-\sigma_q} ] J^{[\nu-1]} 
   e^{ -\hf \tbe^2 \sigma_p\sigma_q \tG(a)}
\nonumber\\
&=&
  \sum_{\nu=1}^\infty {1\over (\nu-2)!}
\int_{D_0}{d^d x_1\over a^d}\cdots  
 \int_{D_{\nu-2}'}{d^d x_{\nu-1}\over a^d}
 \sum_{\sigma_1\not=0}w_{\sigma_1}\cdots 
 \sum_{\sigma_{\nu-2}\not=0}w_{\sigma_{\nu-2}}
 \sum_{\sigma_t=\sigma_{\nu-1}\not=0}w_{\sigma_{\nu -1}}
  J^{[\nu-1]}\,.
\end{eqnarray}
A calculation of this expression finally yields
\begin{equation}
  \delta_f w_\sigma = \hf \Omega_d {\delta a\over a} 
  \sum_{\sigma'\not=\sigma} w_{\sigma'}w_{\sigma-\sigma'}
 \exp\left[ -\hf \tbe^2 \sigma'(\sigma-\sigma') \tG(a)\right]\,.
\end{equation}
Thus, fusion generates a renormalization of the fugacities. In the 
manipulations above, we considered all charge configurations  as
identical if they can be obtained from a given one 
via permutations only. 

\item {\underline {Annihilation.}}

The charges $\sigma_p$ and $\sigma_q=-\sigma_p$ fuse into a neutral 
particle when the inequality $a<|x_p-x_q|<a+\delta a$
holds for their distance. This is called annihilation, 
since both charges vanish from the gas. Effectively,
the term $Z^{[\nu]}$ then yields a contribution to 
$Z^{[\nu-2]}$. In Eqs. \eq{faz} and \eq{jexp} one has to keep the 
terms quadratic in the separation coordinate $s$ (those of the order 
$a^{2(d-2)}$ in the 
exponent of $J^{[\nu]}$) in order to find all terms of the order 
${\cal O} (\delta a/ a)$. Instead of the integral \eq{fusin},
we now get 
\begin{eqnarray}\label{anin}
\lefteqn{\int_{D_{\nu-2}'}{d^d t\over a^d}\int_{a<|s|<a+\delta a} {d^d s\over a^d}
   J^{[\nu]}
=
   J^{[\nu-2]} \exp\left[ +\hf \tbe^2 \sigma_p^2 \tG(a)\right]}
\nonumber\\[2ex]
&& \times 
\int_{D_{\nu-2}'}{d^d t\over a^d}\int_{a<|s|<a+\delta a} {d^d s\over a^d}
\exp\left[ - \tbe^2 \sum_{j\not=p,q} \sigma_j \sigma_p s_\alpha
\partial_\alpha^t  \tG_{x_j,t} +\ord{s^3} \right]
   \nonumber\\[2ex]
&&\approx
   J^{[\nu-2]}  \exp\left[ \hf \tbe^2 \sigma_p^2 \tG(a)\right]
 \int_{D_{\nu-2}'}{d^d t\over a^d}\int_{a<|s|<a+\delta a} {d^d s\over a^d}
\nonumber\\[2ex]
&& 
\biggl[ 1
 - \tbe^2 \sum_{j\not=p,q} \sigma_j \sigma_p s_\alpha
\partial_\alpha^t  \tG_{x_j,t}
   + \hf \tbe^4  \sigma_p^2 \sum_{j\not=p,q} \sum_{i\not=p,q} \sigma_j\sigma_i
s_\alpha s_\beta \, \partial_\alpha^t  \tG_{x_j,t}\partial_\beta^t  \tG_{x_i,t}
\biggr].
\end{eqnarray}
Here the linear term and the off-diagonal part of the matrix $s_\alpha s_\beta$
 give vanishing contributions to the integral
over the annulus, so that one gets
\begin{eqnarray}
\int_{D_{\nu-2}'}{d^d t\over a^d}\int_{a<|s|<a+\delta a} {d^d s\over a^d}
   J^{[\nu]}
&\approx&
   J^{[\nu-2]}  \exp\left[ \hf \tbe^2 \sigma_p^2 \tG(a) \right]\,
\int_{D_{\nu-2}'}{d^d t\over a^d}\int_{a<|s|<a+\delta a} {d^d s\over a^d}
\nonumber\\
&& \times \biggl[ 1
   + {1\over 2d} s^2\tbe^4  \sigma_p^2 \, \sum_{j\not=p,q}
 \sum_{i\not=p,q} \sigma_j\sigma_i
\partial_\alpha^t  \tG_{x_j,t}\partial_\alpha^t  \tG_{x_i,t}
\biggr].
\end{eqnarray}
We now change the order of integration 
so that the annihilating charges are the ``last ones in the 
queue'' and integrate by parts in the second term,
with the result
\begin{eqnarray}\label{anin2}
\int_{D_{\nu-2}'}{d^d t\over a^d}\int_{a<|s|<a+\delta a} {d^d s\over a^d}
   J^{[\nu]}
&\approx&
 J^{[\nu-2]}  \exp\left[ \hf \tbe^2 \sigma_p^2 \tG(a)\right] 
\Omega_d {\delta a\over a} 
\nonumber\\[2ex]
& & \biggl\{ \int_{D_{\nu-2}'}{d^d t\over a^d} 1
  + {1\over 2d}\tbe^4 a^{2-d}  \sigma_p^2 \sum_{j,i=1}^{\nu -2}
    \sigma_j\sigma_i
\nonumber\\
&&
\biggl( \oint_{\partial D_0} -\sum_{l=1}^{\nu-2} \oint_{\partial K_l}
  \biggr) d\sigma_\alpha \tG_{x_j,t}\partial_\alpha^t \tG_{x_i,t} 
\nonumber\\
&& -   {1\over 2d}\tbe^4 a^{2-d}  \sigma_p^2 \sum_{j,i=1}^{\nu -2}
   \int_{D_{\nu-2}'}{d^d t\over a^d} \tG_{x_j,t} \partial^t\partial^t 
\tG_{x_i,t} 
\biggr\},
\end{eqnarray}
where the last integral vanishes since $\partial\partial G(x)\sim \delta (x)$
and the point $t=x_i$ is excluded from the region of integration. The first 
term on the r.h.s. can be included in the renormalization of the free energy
which, however, we do not consider here. Therefore, the only 
nonvanishing terms originate 
from the surface integrals on the r.h.s. of Eq. \eq{anin2}.
They should be considered separately. The integral over
the surface of the domain $D_0$ vanishes as the integrand goes sufficiently
fast to zero for $|t|\to \infty$. The integrand of the integral
over the surface $\partial K_l$ of the sphere centered at $x_l$ contains 
the derivative 
\begin{equation}
 \partial_\alpha^t \tG_{x_i,t}  =
 {(t -x_i)_\alpha \over |t-x_i|^2}  \biggl\{ \begin{array}{lll}
   - c_2 ,&\mbox { for } &d=2 \,,\\
\hf (d-2) \tG_{x_i,t} ,&\mbox { for } &d\geq3\,.
     \end{array} 
\end{equation}
Using the outer normal $n_\alpha$ of the surface $\partial K_l$ and writing
$t= x_l+an$, we obtain:
\begin{equation}\label{kint}
\oint_{\partial K_l} d\sigma_\alpha 
  \tG_{x_j,t} \, \partial_\alpha^t \tG_{x_i,t}=
 \oint_{\partial K_l} df \, n_\alpha
   \tG( x_l+an-x_j) \, \partial_\alpha^l\tG(x_l+an-x_i) =: I\,.
\end{equation}
Here we have to distinguish the following particular cases:
\begin{enumerate}
\item {For \em $l\not=i,j$:}
\begin{eqnarray}\label{lnotij}
I &=& \oint_{\partial K_l} df n_\alpha \left[
 \tG( x_l-x_j)\partial_\alpha^l\tG(x_l-x_i) \right.
\nonumber\\[2ex]
& & \left. +an_\beta \partial_\beta^l\tG( x_l-x_j)\partial_\alpha^l\tG(x_l-x_i)
    +an_\beta \tG( x_l-x_j)\partial_\beta^l\partial_\alpha^l\tG(x_l-x_i)
\right]\,.
\end{eqnarray}
Here, the first term vanishes. The last term vanishes too as it is
 proportional 
to $\delta(x_l-x_i)$, but we have explicitly $x_l\not=x_i$.
The second term can be evaluated as
\begin{eqnarray}
\approx \Omega_d {a^d\over d} \partial_\alpha^l\tG( x_l-x_j)
\partial_\alpha^l\tG(x_l-x_i) 
=
\Omega_d {a^d\over d} { ( x_{l,j})_\alpha(x_{l,i})_\alpha
   \over | x_{l,j}|^2|x_{l,i}|^2}
      \biggl\{ \begin{array}{lll}
 c_2^2,& \mbox{ for }&d=2\,,\\
 {1\over 4}(d-2)^2 \tG( x_{l,j})\tG(x_{l,i}),& \mbox{ for }&d\ge 3\,,
\end{array}
\end{eqnarray}
with $x_{l,j}= x_l-x_j$.
\item  {For \em  $l=j\not=i$:}
\begin{eqnarray}
I &= &\oint_{\partial K_l} d\sigma_\alpha 
  \tG_{x_j,t} \, \partial_\alpha^t \tG_{x_i,t} = 
\oint_{\partial K_l} df \, n_\alpha \, \tG( an)\partial_\alpha^l\tG(x_l+an-x_i)
\nonumber\\
& \approx & \oint_{\partial K_l} df \, n_\alpha \,
   \tG( an) [  \partial_\alpha^l\tG(x_{l,i}) 
     + an_\beta \, \partial_\alpha^l\partial_\beta^l \tG(x_{l,i}) ]
=0,
\end{eqnarray}
where the first term vanishes due to isotropy and the 
second one vanishes, too, for the 
same reason as the last term on the r.h.s. of Eq. \eq{lnotij}.
\item {For \em  $l=i\not=j$:}
\begin{eqnarray}
I &= &\oint_{\partial K_l} d\sigma_\alpha 
  \tG_{x_j,t}\partial_\alpha^t \tG_{x_i,t}
\nonumber\\[2ex]
&=& \oint_{\partial K_l} df n_\alpha
\left.   \tG( x_l+an-x_j)\partial_\alpha^y\tG(y)\right|_{y=an}
\nonumber\\[2ex]
&\approx&
   \oint_{\partial K_l} df n_\alpha
\left[  \tG( x_{l,j})
 +an_\beta \partial_\beta^l  \tG( x_{l,j}) \right]
 \left.\partial_\alpha^y\tG(y)\right|_{y=an}
\nonumber\\[2ex]
&\approx&
   \oint_{\partial K_l}{ df\over a}  \tG( x_{l,j})
    \biggl\{ \begin{array}{lll}
  -c_2 ,& \mbox{ for } & d=2\,,\cr
  \hf (d-2) \tG(an),& \mbox{ for } & d\ge 3 \,,  \end{array}
\nonumber\\[2ex]
&\approx&
  \Omega_d a^{d-2} \tG( x_{l,j}) \biggl\{ \begin{array}{lll}
 - c_2 ,& \mbox{ for } & d=2\,,\cr
  \hf (d-2) \tG(a) ,& \mbox{ for } & d\ge 3\,.  \end{array}
\end{eqnarray}

\item  {For \em  $l=i=j$:}
\begin{eqnarray}
I &=& \oint_{\partial K_l} d\sigma_\alpha 
  \tG_{x_j,t}\partial_\alpha^t \tG_{x_i,t}
\nonumber\\[2ex]
&=& \oint_{\partial K_l} df n_\alpha
   \tG( an)\left. \partial_\alpha^y\tG(y)\right|_{y=an}
\nonumber\\[2ex]
&=& \biggl\{ 
\begin{array}{lll}
0,&\mbox{ for } & d=2\,,\cr
\hf \Omega_d a^{d-2} (d-2)\tG^2(a),&\mbox{ for }& d\ge3\,.
\end{array}
\end{eqnarray}
\end{enumerate}
By employing all the above results, we are now finally 
in a position to rewrite the r.h.s. of Eq. \eq{anin2}:
\begin{eqnarray}
\lefteqn{J^{[\nu-2]} \, \exp\left[ \hf \tbe^2 \sigma_p^2 \tG(a) \right]\,
\Omega_d{\delta a\over a} {1\over a^d}
 \Biggl[  TV- (\nu-2) {\Omega_d a^d\over d}}
\nonumber\\
&&
 - {1\over 2d}\tbe^4 a^{2-d}  \sigma_p^2
\biggl[ \sum_{j,i=1}^{\nu -2}\sigma_j\sigma_i
  \sum_{l\not=i,j}^{\nu-2}
\Omega_d {a^d\over d} { ( x_{l,j})_\alpha(x_{l,i})_\alpha
   \over | x_{l,j}|^2|x_{l,i}|^2}
      \biggl\{ \begin{array}{l}
 c_2^2\cr
 {1\over 4}(d-2)^2 \tG( x_{l,j})\tG(x_{l,i})
\end{array}\biggr\}
\nonumber\\
&&
 +  \sum_{j}^{\nu -2}\sigma_j\sum_{l\not=j}^{\nu-2}\sigma_l
 \Omega_d a^{d-2} \tG( x_{l,j}) \biggl\{ \begin{array}{l}
  c_2 \cr
  \hf (d-2) \tG(a)  \end{array}\biggr\}
  +  \sum_{j}^{\nu -2}\sigma_j^2
\hf \Omega_d a^{d-2} (d-2)\tG^2(a)
\biggr]
\Biggr],
\end{eqnarray}
where, again, the first and second rows of the columns stand for $d=2$ 
and $d\ge 3$,
respectively. Multiplying this expression by $w_{\sigma_p}w_{-\sigma_p}$
and summing up for $\sigma_p$, one obtains:
\begin{eqnarray}\label{anin3}
\lefteqn{J^{[\nu-2]}\Omega_d{\delta a\over a}
\Biggl[ {1\over a^d}\biggl( TV- (\nu-2) {\Omega_d a^d\over d} \biggr)
  \sum_{\sigma\not=0} w_\sigma w_{-\sigma} \,
\exp\left[ \hf \tbe^2 \sigma^2 \tG(a)\right]}
\nonumber\\
&&
 - {1\over 2d}\tbe^4 a^{2}  
\biggl[ \sum_{j,i=1}^{\nu -2}\sigma_j\sigma_i
  \sum_{l\not=i,j}^{\nu-2}
\Omega_d {1\over d} { ( x_{l,j})_\alpha(x_{l,i})_\alpha
   \over | x_{l,j}|^2|x_{l,i}|^2}
      \biggl\{ \begin{array}{l}
 c_2^2\cr
 {1\over 4}(d-2)^2 \tG( x_{l,j})\tG(x_{l,i})
\end{array}\biggr\}
\nonumber\\
&&
 +  \sum_{j}^{\nu -2}\sigma_j\sum_{l\not=j}^{\nu-2}\sigma_l
 \Omega_d a^{-2} \tG( x_{l,j}) \biggl\{ \begin{array}{l}
  c_2 \cr
  \hf (d-2) \tG(a)  \end{array}\biggr\}
  +  \sum_{j}^{\nu -2}\sigma_j^2
\hf \Omega_d a^{-2} (d-2)\tG^2(a)
\biggr]
<\sigma^2>
\Biggr]
\end{eqnarray}
with 
\begin{equation}
<\sigma^2>=  \sum_{\sigma\not=0} \sigma^2 w_\sigma w_{-\sigma}
\exp\left[  \hf \tbe^2 \sigma^2 \tG(a) \right].
\end{equation}
With the help of this relation, we can now write down the 
infinitesimal additional contribution to the 
partition function under the blocking transformation,
\begin{eqnarray}
\lefteqn{Z+\delta_a Z=  \biggl( 1 + 
\Omega_d{\delta a\over a}
{TV\over a^d}
  \sum_{\sigma\not=0} w_\sigma w_{-\sigma} 
\exp\left[ \hf \tbe^2 \sigma^2 \tG(a) \right] 
\biggr)
  \sum_{\nu=2}^\infty {1\over \nu!}{\nu!\over (\nu-2)!2!}}
\nonumber\\
&&
 \int_{D_0}{d^d x_1\over a^d}\cdots \int_{D_{\nu-3}'}{d^d x_{\nu-2}\over a^d} 
\biggl[
\prod_{j=1}^{\nu-2}
\sum_{\sigma_j} w_{\sigma_j}
 \biggl(
  1 - {\delta a\over a}
 {\Omega_d^2 \over d}
  \sum_{\sigma\not=0} w_\sigma w_{-\sigma} 
  \exp\left[ \hf \tbe^2 \sigma^2 \tG(a)\right]
\nonumber\\
&& - {d-2\over 4 d}\tbe^4 \, {\delta a\over a} \, \Omega_d^2\,
    \sigma_j^2  \, \tG^2(a) <\sigma^2>
\biggr)
\biggr]
\nonumber\\
&&
J^{[\nu-2]} \biggl( 1 - \tbe^4 { \Omega_d^2\over  d} {\delta a\over a}
  \biggl\{ \begin{array}{l}
 - c_2 \cr
  \hf (d-2) \tG(a)  \end{array}\biggr\}
 {1\over 2} \sum_{j}^{\nu -2}\sum_{l\not=j}^{\nu-2}
  \sigma_j\sigma_l
   \tG( x_{l,j}) <\sigma^2>
\biggr).
\end{eqnarray}
Here, the first term leads to the renormalization of the free energy.
The contribution of the third term on the r.h.s. of Eq. \eq{anin3} 
vanishes that can be seen after partial integration since $x_{l,i}$ 
and $x_{l,j}$ are not zero. The effects of all the other terms merge 
in the modifications of the parameters of the original partition function:
\begin{subequations}
\begin{eqnarray}
\delta_a \tbe^2 &=& \tbe^4 { \Omega_d^2\over  d} {\delta a\over a}
  \biggl\{ \begin{array}{l}
  -c_2 \cr
  \hf (d-2) \tG(a)  \end{array}\biggr\}
  <\sigma^2> ,
\nonumber\\
 \delta_a w_\sigma &=&
 - {\delta a\over a} w_\sigma
 {\Omega_d^2 \over d}
  \sum_{\sigma'\not=0} w_{\sigma'} w_{-\sigma'} \,
    \exp\left[ \hf \tbe^2 {\sigma'}^{2} \tG(a)\right]
 - {d-2\over 4 d}\tbe^4  w_\sigma  {\delta a\over a}\Omega_d^2
    \sigma^2   \tG^2(a) <{\sigma'}^{2}>.
\end{eqnarray}
\end{subequations}
\end{enumerate}
As we sum the various contributions to the changes of the parameters, 
we finally arrive at the flow equation \eq{rswcg}, which completes 
its derivation.

\newpage


%
%

%
%

%
%

%
%

%
%

%
%

%
%

%
%


\newpage


\begin{thebibliography}{100}
\bibitem{kt} J. M. Kosterlitz, D. J. Thouless, J. Phys. {\bf C6} (1973) 118;
        J. M. Kosterlitz, J. Phys. {\bf C7} (1974) 1046.
\bibitem{col} S. Coleman, Phy. Rev. {\bf D11} (1975) 3424.
\bibitem{jkkn} J. V. Jose, L. P. Kadanoff, S. Kirkpatrick, D. R. Nelson,
        Phys. Rev. {\bf B16} (1977) 1217.
\bibitem{wieg} P. B. Wiegmann, J. Phys. {\bf C11} (1978) 1583.
\bibitem{samu} S. Samuel, Phys. Rev. {\bf D18} (1978) 1916.
\bibitem{agg} D. Amit, Y. Y. Goldschmidt, G. Grinstein, 
        J. Phys. {\bf A13} (1980) 585.
\bibitem{Nienhuis} B. Nienhuis in 
        {\em Phase Transitions and Critical Phenomena}, Vol. 11 ed. 
        by C. Domb, J.L. Lebowitz (Academic Press, London, 1987), 1-53.
\bibitem{kerson} K. Huang, J. Polonyi, Int. J. Mod. Phys. {\bf A6} 
        (1991) 409.
\bibitem{creswick} R. J. Creswick, H. A. Farach, C. P. Poole, Jr.
        Introduction to RG methods in Physics, (Wiley, New York, 1992)
\bibitem{gulacsi} Zs. Gul\'acsi, M. Gul\'acsi, Adv. in Phys. {\bf 47} 
        (1998) 1-89.
\bibitem{wet} G. v. Gersdorff, C. Wetterich, Phys. Rev. {\bf B64} (2001)
        054513. 
\bibitem{per} I. N\'andori, J. Polonyi, K. Sailer, Phys. Rev. {\bf D 63}  
        (2001) 045022.
\bibitem{coulomb} I. N\'andori, J. Polonyi, K. Sailer, Phil. Mag. {\bf B 81} 
        (2001) 1615. 
\bibitem{jpg}  I. N\'andori,  K. Sailer, U. D. Jentschura, G. Soff, J. Phys. 
        {\bf G 28} (2002) 607.
\bibitem{fertig} H. A. Fertig, Kingshuk Majumdar, cond-mat/0302012.
\bibitem{deconf} J. Polonyi, in {\em Quark Gluon Plasma},
        World Scientific, 1990, R. Hwa, ed.;
        K. Johnson, L. Lellouch, J. Polonyi, Nucl. Phys. {\bf B367} (1991) 675;
        Act. Phys. Hung. {\bf 2} (1995) 123.
\bibitem{savit} R. Savit, Rev. Mod. Phys. {\bf 52}, (1980) 453.
\bibitem{wh} F. J. Wegner, A. Houghton, Phys. Rev. {\bf A8} (1973) 401.
\bibitem{nelson} D. Nelson and D. Fisher, Phys. Rev. {\bf B16} (1977) 4945.
\bibitem{shenoy} S. R. Shenoy, Phys. Rev. {\bf B40} (1989) 5056. 
\bibitem{spaceshuttle} M. Campostrini, M. Hasenbusch, A. Pelisetto, P. Rossi,
        E. Vicari, Phys. Rev. {\bf B63} (2001) 214503
\bibitem{choi} M. S. Choi, S. I. Lee, Phys. Rev. {\bf B51} (1995) 6680.
\bibitem{shenoy2} S. R. Shenoy, B. Chattapadhyay, Phys. Rev. {\bf B51} (1995) 9129.
\bibitem{lsc} B. Sas {\em et al.}, Phys. Rev. {\bf B61} (2000) 9118;
        I. Pethes {\em et al.}, Synthetic Metals {\bf 120} (2001) 1013; 
        F. Portier {\em et al.}, Phys. Rev. {\bf B66} (2002) 140511;    
        K. Vad {\em et al.}, 2003, to be published.     
\bibitem{conv} R. J. Rivers, {\em Path Integral Methods in 
        Quantum Field Theory}, (University Press, Cambridge, 1987).
\bibitem{tree} J. Alexandre, J. Polonyi, Phys. Lett. {\bf B 445} (1999) 351.
\bibitem{morris} T.R. Morris, J.F. Tighe, JHEP {\bf 9908} (1999) 007,
        T.R. Morris, M.D. Turner, Nucl. Phys. {\bf B509} (1998) 637,
        T.R. Morris, Nucl. Phys. {\bf B495} (1997) 477,
        ibid. Int. J. Mod. Phys. {\bf B12} (1998) 1343.
\bibitem{mermin} N.D. Mermin, H. Wagner, Phys. Rev. Lett. {\bf 17} (1966)
        1133.
\bibitem{wilson} K. G. Wilson, J. Kogut, Phys. Rep. {\bf C12} (1974) 77; 
        K. G. Wilson, Rev. Mod. Phys. {\bf 47} (1975) 773; 
        Rev. Mod. Phys. {\bf 55} (1983) 583.    
\bibitem{polc} J. Polchinski, Nucl. Phys. {\bf B231} (1984) 269.
\bibitem{comellas} J. Comellas, Nucl. Phys. {\bf B509} (1998) 662 
\bibitem{ball} R. D. Ball, P. E. Haagensen, J. I. Latorre, E. Moreno, 
        Phys. Lett. {\bf B347} (1995) 80 
\bibitem{kornelsajat} J. Polonyi, K. Sailer, cond-mat/0108179;
        Phys. Rev. {\bf B66} (2002) 155113.
\bibitem{O(N)} J. Comellas, A. Travesset, Nucl. Phys. {\bf B 498} (1997) 2411.
\bibitem{senben} S. Liao, J. Polonyi, Ann. Phys. {\bf 222} (1993) 122.
\bibitem{hh} A. Hasenfratz, P. Hasenfratz, Nucl. Phys. {\bf B270} (1986) 687.
\bibitem{janosRG} J. Polonyi, Lectures on Functional Renormalization 
        Group Method, hep-th/0110026.
\bibitem{ber} C. Bagnuis, C. Bervillier, Phys. Rept. {\bf 348} (2001) 91
\bibitem{lowd}  J. M. Carmona, J. Polonyi, A. Taranc\'on, 
        Phys. Rev. {\bf D61} (2000) 085018 
\end{thebibliography}
\end{document}